\documentclass[preprint,epsbox]{jpsj2}
\def\runtitle{Ordering of the Antiferromagnetic Heisenberg Model 
on a Pyroclhore Slab}
\def\runauthor{Takuya \textsc{Arimori} and Hikaru \textsc{Kawamura}}
\title{
Ordering of the Antiferromagnetic Heisenberg Model
\\ on a Pyrochlore Slab}
\author{
Takuya Arimori and Hikaru Kawamura
}
\inst{
Department of Earth and Space Science, Graduate School of Science, Osaka
University,\\
Toyonaka 560-0043
}
\recdate{July 30}
\abst{
Ordering of the geometrically frustrated two-dimensional
Heisenberg antiferromagnet on a pyrochlore slab
is studied by
Monte Carlo simulations. The model is expected to serve as a
reference system of SrCrGaO compound studied extensively.
In sharp contrast to the kagom\'e Heisenberg antiferromagnet,
the model
exhibits locally non-coplanar  spin
structures at low temperatures,
bearing nontrivial chiral degrees of freedom.
We find that
under certain conditions the model exhibits a novel
Kosterlitz-Thouless-type transition at a finite temperature
associated with these chiral
degrees of freedom. Implications to experiments are discussed.
}
\kword{frustration, chirality, pyrochlore slab, Monte Carlo simulation,
kagom\'e lattice
}
\begin{document}
\sloppy
\maketitle

\section{Introduction}

Magnetic ordering of geometrically frustrated antiferromagnets (AFs)
has attracted continual interest of researchers in magnetism\cite{Diep}.
In geometrically frustrated
AFs,  spins usually
sit on lattices made up of triangles or tetrahedra
as elementary units, and interact antiferromagnetically
with their neighboring spins. Intrinsic inability to 
simultaneously satisfy all antiferromagnetic 
nearest-neighbor interactions on a triangle or on a
tetrahedron necessarily leads to macroscopic frustration. This makes the
spin ordering on these lattices a highly nontrivial issue.
Triangles- or tetrahedra-based lattices
might be classified into two categories: One is the tightly
coupled lattice consisting of {\it edge\/}-sharing
triangles or tetrahedra,
and the other is the loosely-coupled lattice  consisting of
{\it corner\/}-sharing triangles or tetrahedra. Examples of the former are
the triangular lattice in two dimensions (2D)
and the stacked-triangular lattice in three dimensions (3D),
while those
of the latter are the 2D kagom\'e lattice and the 3D pyrochlore lattice.
In earlier studies, emphasis was put on the former
category\cite{Collins,Kawareview}. These studies
revealed a variety of interesting ordering phenomena not encountered in
standard unfrustrated magnets, {\it e.g.\/},  novel universality classes,
exotic phase transition such as chiral transition and new type of
topological phase transition {\it etc\/}.

Recently, interest has been
focused more on the latter category, {\it i.e.\/}, the 2D
kagom\'e and 3D pyrochlore AFs\cite{Ramirezreview,Schifferreview,Harrisreview}.
Due to the looser coupling among
the frustrating units,
these systems often
remain paramagnetic down to very low temperatures
without any magnetic ordering. Indeed, various theoretical studies on
the 2D kagom\'e and 3D pyrochlore AFs have revealed that these systems
remain paramagnetic down to zero temperature without any
phase transition\cite{Chalker,Reimers1,Reimers2,Moessner}.
Experimentally, however,
many of the geometrically frustrated magnets, which
are regarded as typical
kagom\'e or pyrochlore AFs, exhibit a
phase transition at a low but finite temperature,
quite often a spin-glass (SG)-like freezing
transition\cite{Ramirezreview,Schifferreview,Harrisreview}.

One of the best studied geometrically frustrated AFs is the
$S=3/2$ Heisenberg kagom\'e AF
SrCrGaO (SCGO)\cite{Ramirezreview,Schifferreview}.
Experimentally, this material exhibits a
SG-like transition at a finite temperature $T=T_f$ as in many other
geometrically frustrated AFs, although $T_f$
is considerably
lower than the Curie-Weiss temperature of this material due to the strong
geometrical
frustration\cite{Ramirez,Broholm}.
In spite of extensive experimental and theoretical efforts,
the true nature of this SG-like transition of SCGO
has remained elusive.
Extensive Monte Carlo (MC) simulations performed for
the 2D kagom\'e Heisenberg AF have failed to reproduce the SG-like
transition as experimentally observed in SCGO, suggesting that the modeling
of SCGO as a pure  kagom\'e AF might be inadequate in capturing
some essential aspects of this material.

Although SCGO has
been regarded for some time as a typical model compound of the
2D kagom\'e AF,
the underlying
lattice structure is in fact not of a pure (single-layer)
kagom\'e lattice, but rather, of a
kagom\'e sandwich, or a
``pyrochlore slab''\cite{Ramirez,Broholm,Ramirezreview,Schifferreview}.
The structure of the
lattice is illustrated in Fig.1(a):
It consists of two
2D kagom\'e layers
which sandwiches the sparse triangular
layer in between.
Note that this lattice is obtained by slicing the 3D pyrochlore
lattice along the (111) direction into the slab geometry.
Inelastic neutron-scattering measurements have indicated that
the neighboring slabs are
magnetically well separated along the $c$-axis\cite{Lee}.
Hence, in modeling SCGO,
one may safely neglect the inter-slab interaction
and consider the 2D Heisenberg model on a pyrochlore-slab lattice.

The  purpose of the present paper is to study the ordering properties
of the antiferromagnetic classical Heisenberg model on the pyrochlore-slab
lattice by means of
MC simulations, and to examine whether some new
features which are different from those
of the well-studied pure kagom\'e Heisenberg AF would arise,
possibly due to the tetrahedron-based structure
of this lattice. In particular, we pay attention to the possible ``chiral''
properties of the model. ``Chirality'' is  a multi-spin quantity
representing the sense or handedness of the local non-coplanar spin structures
induced by spin frustration. It is defined for
three neighboring Heisenberg spins as a pseudo-scalar, 
$\chi =\mbox{\boldmath $S$}_1\cdot\mbox{\boldmath $S$}_2
\times \mbox{\boldmath $S$}_3$, so as to give a nonzero value if the the three
spins make non-coplanar configurations but vanish  otherwise.
This type of chirality is sometimes called ``scalar chirality'', which
is distinct from the ``vector chirality'' often used in the literature,
defined for two neighboring Heisenberg spins by an axial vector
$\mbox{\boldmath $S$}_1
\times \mbox{\boldmath $S$}_2$\cite{Kawamiya}.
In the case of the pure kagom\'e Heisenberg AF, it has been known that
the spin structure selected at low temperatures
is a coplanar one with the vanishing scalar
chirality\cite{Chalker,Reimers2}.
In sharp contrast, we shall show below that in the case of the
pyrochlore-slab Heisenberg AF
the spin structure stabilized at low temperatures is a
{\it non-coplanar\/}
one sustaining nontrivial
chiral degrees of freedom.  Depending on the parameter values of our
model Hamiltonian, these nontrivial chiral degrees of freedom
are found to exhibit novel thermodynamic
phase transition at a finite temperature without accompanying the
order of Heisenberg spins. Such chiral phase transition, unexpected so far,
does not occur in the kagom\'e Heisenberg AF.

In section 2, we introduce our model and explain some of the
details of our numerical method. Our model Hamiltonian
possesses, within the kagom\'e layers,
the antiferromagnetic nearest-neighbor (nn) coupling $J_1>0$ and
next-nearest-neighbor (nnn) coupling $J_2$ of either sign, while it possesses,
between the kagom\'e layers and the triangular layer,
the antiferromagnetic nn coupling $J'>0$.
Various physical
quantities calculated in MC simulations are defined in section 3.
The results of our MC simulations are presented in section 4.
The cases of vanishing, antiferromagnetic and ferromagnetic
nnn interactions, {\it i.e.\/}, the cases of
$J_2=0,\ J_2>0,\ J_2<0$, are dealt with in section 4.1, 4.2 and 4.3,
respectively. In all cases studied, the system is found to
sustain  nonzero local chiralities, in sharp contrast to the
pure kagom\'e Heisenberg AF. Furthermore,
in the particular case of the antiferromagnetic
nnn coupling $J_2>0$, we find that the
model exhibits a Kosterlitz-Thouless-type transition
associated with the chiralities.
Section 5 is devoted to summary and
discussion.  Implication to experiments is briefly discussed.

\section{The Model and the Method}

The model we consider is the isotropic classical Heisenberg
model on a pyrochlore-slab (or kagom\'e-sandwich) lattice.
The pyrochlore-slab lattice is illustrated in Fig.1(a).
It consists of two
2D kagom\'e layers of lattice spacing $d$
which sandwiches the sparse triangular
layer of lattice spacing $2d$ in between.
The unit cell of the lattice may
be taken as two corner-sharing tetrahedra containing seven
sites numbered from 1 to 7, which
is illustrated  by the solid lines in Fig.1(a).
Among these seven sites
belonging to a unit cell,
the lower threes forming the equilateral triangle
(the sites 1, 2 and 3 in Fig.1(a)) are parts of
the lower kagom\'e layer,
the upper threes
forming the equilateral triangle (the sites 5, 6 and 7)
are parts of the upper kagom\'e layer,
and the one in the middle (the site 4)
is a part of the sparse triangular layer.
These unit cells containing seven sites are arranged forming the 2D
triangular lattice of spacing $2d$. 
In Fig.1(b), we show the lower kagom\'e layer,
which consists of only the sites
1, 2 and 3. Note that the upward triangle
in Fig.1(b)
corresponds to the bottom plane of tetrahedron with an apical site 4 on
top of it, while the downward triangle is not a part of any tetrahedron.
These
upward triangles are further grouped into three types, denoted
$A, B$ and $C$, each forming triangular sublattices of spacing
$2\sqrt 3d$ (recall here that the triangular lattice can be
decomposed into three
inter-penetrating triangular sublattices). We shall use such
representation of the lattice later.

Our Hamiltonian is given by
\begin{equation}
{\cal H} = J_1\sum_{\langle ij\rangle }^{n.n.}
\mbox{\boldmath $S$}_i \cdot \mbox{\boldmath $S$}_j
+J_2\sum_{\langle kl\rangle }^{n.n.n.}
\mbox{\boldmath $S$}_k \cdot \mbox{\boldmath $S$}_l
+J^{'}\sum_{\langle mn\rangle }^{}
\mbox{\boldmath $S$}_m \cdot \mbox{\boldmath $S$}_n,
\end{equation}
where $J_1>0$ is the antiferromagnetic
nn interaction on the two kagom\'e layers,
$J_2$ is the nnn interaction on the kagom\'e layers,
and $J^{'}(>0)$ is the antiferromagnetic nn interaction
between the kagom\'e layers and the triangular layer: See Fig.1(a).
The first and second sums in eq.(1)
are taken over all nn and
nnn pairs on the two kagom\'e layers,  while
the third sum is taken over all nn pairs linking
the triangular layer and  the two kagom\'e layers.
The variable $\mbox{\boldmath $S$}_i$
is a three-component unit vector
representing a classical Heisenberg spin at the $i$-th site.

In order to investigate the thermodynamic properties of this model,
we perform the standard
heat-bath MC simulations.
Since the present model is a highly frustrated model possessing
many degenerate states, possibly leading to very slow relaxation,
we combine the heat-bath method
with the temperature-exchange technique to facilitate efficient
thermalization\cite{HN}.
Simulations are made for a pyrochlore-slab lattice with
$N=7\times L\times L$ spins
with  $L=$6, 12, 18, 24 and 30, where the number seven here
represents the number of spins per unit cell.
Periodic boundary conditions are employed.

Typically, initial $2\times 10^6$ Monte Carlo steps per spin
(MCS) are discarded for thermalization,
and the following $1.8\times 10^7$ MCS are
used to calculate various physical quantities.
The latter $1.8\times 10^7$ MCS is divided into 5 bins, and the
error bars are estimated from the standard deviation of the data sets
taken for these 5 bins.

Although our model is a regular one without any quenched randomness,
it turns out to be a hard-relaxing system
due to its severe frustration, exhibiting 
very slow relaxation at low temperatures similar to the ones
encountered in spin glasses.
Therefore, we pay special attention to be sure that the system is
fully thermalized.
Equilibration is checked by the following procedures:
First, we monitor the system to travel back and forth
many times during the
the temperature-exchange process (typically more than 10 times)
between the maximum and minimum temperature points, and check
at the same time that the relaxation
due to the standard heat-bath updating
is reasonably fast at the highest temperature,
whose relaxation time is of order $10^2$ MCS.
This guarantees that significantly different parts of
the phase space is sampled in each ``cycle'' of the temperature-exchange
process.
Second, we monitor the stability of the calculated physical quantities to
check that they remain stable during at least  
three times longer MC period.

\section{Physical Quantities}

In this section, we
define various physical quantities calculated in our
simulations below.
Energy, specific heat and uniform magnetic susceptibility
are defined and calculated in the standard way.
We calculate in addition
the nonlinear magnetic susceptibility
$\chi_2$, according to the relation,
\begin{equation}
\chi_2=\frac{1}{6N k_{\rm B}T}
(\langle M_z^4 \rangle -4\langle M_z\rangle\langle M_z^3 \rangle
-3\langle M_z^2 \rangle^2+12\langle M_z^2 \rangle\langle M_z \rangle^2
-6\langle M_z \rangle^4),
\end{equation}
where $M_z$ is the $z$-component of the total magnetization of the system,
and $\langle \cdots \rangle$ represents the thermal average.
One generally expects that there should be
no long-range order (LRO) of Heisenberg spin, 
nor a finite-temperature
transition occurring in its spin sector, 
in a fully isotropic Heisenberg model in
two spatial dimensions like our model.
Nevertheless, in order to
probe the
possible development of the spin short-range order (SRO),
we follow the previous works on the pure kagom\'e AF
and calculate the following two Fourier modes of spin order,
the $q=0$ mode and the $\sqrt 3\times \sqrt 3$ mode, each
being 
the representative ordering mode of the kagom\'e 
lattice\cite{Chalker,Reimers2}.
The Fourier magnetization associated with the $q=0$ mode is defined by
\begin{equation}
m_0=\langle|{\mbox{\boldmath $m$}}_{0}|^2\rangle^{1/2},
\end{equation}
\begin{equation}
{\mbox{\boldmath $m$}}_{0}=\frac{\sqrt{2}}{3N_{s}}\sum_{i, \alpha}
{\mbox{\boldmath $S$}}_i^{\alpha}
{\rm exp}({\rm i} \phi_{\alpha}),
\end{equation}
where $N_s=L\times L$ denotes the total number of unit cells, 
and $(\phi_1,\phi_2,\phi_3)=(0,2\pi/3,4\pi/3)$. The summation over $i$ is
taken over $N_s$ unit cells
and that over $1\leq \alpha \leq 3$ is taken over
three sites in a unit cell lying on the lower kagom\'e layer.
Note that $m_0$ gives unity
when the spin configuration is in the $q=0$ state.
The Fourier magnetization associated with the
$\sqrt 3\times \sqrt 3$ mode is defined by
\begin{equation}
m_{\sqrt 3}=\langle|{\mbox{\boldmath $m$}}_{\sqrt 3}|^2\rangle^{1/2},
\end{equation}
\begin{equation}
{\mbox{\boldmath $m$}}_{\sqrt{3}}=\frac{\sqrt{2}}{3N_{s}}(\sum_{i\in{A}
,\alpha}
{\mbox{\boldmath $S$}}_i^{\alpha}{\rm exp}({\rm i} \phi_{\alpha}^
{A})+\sum_{i\in{B},{\alpha}}{\mbox{\boldmath $S$}}_
i^{\alpha}{\rm exp}({\rm i} \phi_\alpha^{B})+\sum_{i\in{C},
{\alpha}}{\mbox{\boldmath $S$}}_i^{\alpha}{\rm exp}
({\rm i} \phi_{\alpha}^{C})),
\end{equation}
where $(\phi_1^{A},\phi_2^{A},\phi_3^{A})=(0,2\pi/3,4\pi/3)$,
$(\phi_1^{B},\phi_2^{B},\phi_3^{B})=(2\pi/3,4\pi/3,0)$,
$(\phi_1^{C},\phi_2^{C},\phi_3^{C})=(4\pi/3,0,2\pi/3)$, while 
$A$, $B$ and $C$ denote the three triangular sublattices shown in
Fig.1(b). Note that
$m_{\sqrt{3}}$ gives unity
when the spin configuration is in the $\sqrt{3} \times \sqrt{3}$ state.
In our definition of ${\mbox{\boldmath $m$}}_{\sqrt{3}}$ and
${\mbox{\boldmath $m$}}_0$ above, we have implicitly assumed
that the antiferromagnetic $J_1$ tends to align any
three spins at each
elementary triangle into the $120^\circ$ spin structure
in which neighboring spins make an angle equal to $2\pi/3$
with each other.

One can also define the Binder ratio associated with these ferrimagnetic order
parameters. For example, the Binder ratio associated with the $q=0$ mode may
be defined by
\begin{equation}
g_{0}=4-3\frac{\langle |{\mbox{\boldmath $m$}}_{0}|^4\rangle}
{\langle |{\mbox{\boldmath $m$}}_{0}|^2\rangle^2}.
\end{equation}
Here, $g_{0}$ is defined so that in the
thermodynamic limit it vanishes in the high-temperature phase while
it gives unity in the non-degenerate ordered phase. 
We have used 
the fact that
the order parameter considered here has six independent components.
    
As mentioned, we are particularly interested
in the ordering behavior of the chirality.
Generally, the local chirality may be defined
for three neighboring Heisenberg spins.
Here, we calculate the local chirality
for the three spins located at each upward triangle
on the lower kagom\'e layer (the spins 1, 2, 3 in Fig.1(b)),
\begin{equation}
\chi_i=\mbox{\boldmath $S$}_i^1 \cdot (\mbox{\boldmath $S$}_i^2
\times \mbox{\boldmath $S$}_i^3). 
\end{equation}
The upward triangle on the lower kagom\'e layer
corresponds to the bottom plane of tetrahedron.
Chirality may also be defined for other types of triangles as well.
We supplementarily calculate the local chirality 
defined for three Heisenberg
spins on the downward triangle
on the lower kagom\'e layer which is not 
a part of any tetrahedron,
\begin{equation}
\chi_i^{\rm tri}=\mbox{\boldmath $S$}_i^1 \cdot (\mbox{\boldmath $S$}_j^2
\times \mbox{\boldmath $S$}_k^3).
\end{equation}
In order to measure the local non-coplanarity of the spin structure,
we compute the mean local amplitude of these chiralities,
\begin{equation}
\bar \chi^2= \frac{1}{N_{s}}\sum _i
\langle \chi_i^2 \rangle, \ \ \
\bar \chi_{\rm tri}^2= \frac{1}{N_{s}}\sum _i
\langle (\chi_i^{\rm tri})^2 \rangle.
\end{equation}
These quantities vanish for coplanar spin structures, and its magnitude
tells us the extent of the non-coplanarity of the local spin
structures.
Previous studies have revealed that
these quantities tend to vanish at low temperatures
in the pure kagom\'e Heisenberg AF\cite{Chalker,Reimers2}.

In order to detect the ordering of the chirality,
we calculate the ferrimagnetic (or staggered)
chiral order parameter, $m_{\chi f}$, defined by
\begin{equation}
m_{\chi f}^2=(m_{\chi}^A)^2+(m_{\chi}^B)^2+(m_{\chi}^C)^2-m_{\chi}^A
m_{\chi}^B-m_{\chi}^Bm_{\chi}^C-m_{\chi}^Cm_{\chi}^A,
\end{equation}
\begin{equation}
m_{\chi}^A=\frac{3}{N_s}\sum_{i\in A}\chi_i,\ \
m_{\chi}^B=\frac{3}{N_s}\sum_{i\in B}\chi_i,\ \
m_{\chi}^C=\frac{3}{N_s}\sum_{i\in C}\chi_i.
\end{equation}
This quantity gives a nonzero value if the chirality exhibits a ferrimagnetic
order with $\sqrt 3\times \sqrt 3$ periodicity, characterized by
the wavevectors
$\pm \mbox{\boldmath $Q$}=(\pm 4\pi/3d',0)$, where $d'=2d$ is 
the lattice constant of the triangular lattice.

The associated chiral Binder ratio may be defined by
\begin{equation}
g_{\chi f}=2-\frac{\langle m_{\chi f}^4\rangle}
{\langle m_{\chi f}^2\rangle^2}.
\end{equation}
Note that $g_{\chi f}$ is defined here so that in the
thermodynamic limit it vanishes in the high-temperature phase while
it gives unity in the non-degenerate ordered phase. 

Further information on the ferrimagnetic chiral order can be obtained via
the ``phase variable''
$\theta $, defined as follows.
By introducing the ``real'' (or cosine) and the ``imaginary''
(or sine) parts of the Fourier magnetization by
\begin{equation}
m_{\chi}^R=\frac{1}{2}(2m_{\chi}^A-m_{\chi}^B-m_{\chi}^C),
\end{equation}
\begin{equation}
m_{\chi}^I=\frac{\sqrt{3}}{2}(m_{\chi}^C-m_{\chi}^B),
\end{equation}
we define the phase $\theta $  by
\begin{equation}
\theta = {\rm arg}(m_{\chi}^R + {\rm i} m_{\chi}^I). %=\frac{\pi}{3}n.
\end{equation}
If the ordered state is in the ``cosine'' state
with, say, $m_{\chi}^A:m_{\chi}^B:m_{\chi}^C=1:1:-1$, then
the phase $\theta $ becomes a multiple of $\pi/3$, {\it i.e.\/},
it is equal to  $\frac{\pi}{3}n$, with $n$ being an integer. On the other hand,
if the ordered state is in the ``sine state'' (or in
the so-called partial disordered state) with, say,
$m_{\chi}^A:m_{\chi}^B:m_{\chi}^C=1:0:-1$, then
the phase $\theta $ is equal to $\frac{\pi}{3}(n+\frac{1}{2})$.

So far, we have dealt with the possible ferrimagnetic $\sqrt 3\times \sqrt 3$
ordering of the
chirality. In a highly frustrated model like our model,
chiralities might possibly be ordered into more complicated
spatial patterns, or even into spatially random patterns without any spatial
periodicity. In order not to miss such possibilities,
we also calculate the Edwards-Anderson-type chiral order parameter
used in the study of spin glasses, 
``chiral-glass order parameter''\cite{HukuKawa}.
For this,
we first introduce the replica overlap of the scalar chirality $q_\chi$,
by considering two independent systems (``replicas'') described by
the same Hamiltonian (1),
via the relation, 
\begin{equation}
q_{\chi}=\frac{1}{N_{{\rm s}}}\sum_i
\chi_{i}^{(1)}\chi_{i}^{(2)},
\end{equation}
where $\chi_{i}^{(1)}$ and  $\chi_{i}^{(2)}$ represent the chiral
variables defined at the $i$-th upward triangle on the lower kagom\'e
layer
of the replicas 1 and 2, respectively.
In our simulations, we prepare the two replicas 1 and 2 by
running two independent sequences of  systems
in parallel with different spin initial conditions and
different sequences of random numbers.
In terms of this chiral overlap $q_{\chi}$, the chiral-glass
order parameter may be defined by
\begin{equation}
q_{\chi}^{(2)}=\langle q_{\chi}^2\rangle\ \ .
\end{equation}
This quantity gives a nonzero value if there occurs any type of
chirality ordering, either being periodic or random.

\section{Monte Carlo Results}

In this section, we present the results of our MC simulations.
This section is divided into three subsections.
In \S 4.1, we first deal with the case of  vanishing nnn
interaction in the kagom\'e layers, {\it i.e.\/}, the case $J_2=0$.
The cases of the antiferromagnetic and ferromagnetic
nnn interaction in the kagom\'e layers, $J_2>0$ and $J_2<0$,
are dealt with subsequently
in  \S 4.2 and \S 4.3.

\subsection{The case of vanishing next-nearest-neighbor interaction: $J_2=0$}

First, we consider the case in which the interaction in the
kagom\'e layers works only between nearest neighbors, {\it i.e.\/},
the case $J_2=0$.
In Fig.2, we show
the temperature and size dependence of the specific heat per spin.
The specific heat gradually increases with decreasing temperature, without
showing any prominent feature.
In the low-temperature limit $T\rightarrow 0$, the
specific heat tends to
the asymptotic value, $C(T\rightarrow 0)\simeq0.845$,
which is considerably smaller than
the spin-wave value, unity,
expected in the standard classical Heisenberg model.
Similar deviation from the spin-wave value has been known to occur
in the pure kagom\'e Heisenberg AF\cite{Chalker,Reimers2,Moessner},
and is a manifestation of the severe frustration of the model.

In Fig.3, we show
the temperature and size dependence of the linear and nonlinear
susceptibilities per spin. No anomalous behavior is appreciable in these
quantities. The linear susceptibility exhibits only weak temperature
dependence, while the nonlinear susceptibility stays zero within the
error bars.

In Fig.4, we show the the temperature and size dependence of
the ferrimagnetic magnetizations associated with the
$q=0$ and the $\sqrt 3\times \sqrt 3$ modes, $m_{0}$ and $m_{\sqrt 3}$.
One can see from the figures that both $m_{0}$ and $m_{\sqrt 3}$
stay small
even at lower temperatures, and that there is no appreciable selection
between these two modes.
This is in contrast to the behavior
of the pure kagom\'e Heisenberg AF where the  $\sqrt 3\times \sqrt 3$ mode is
selected over the $q=0$ mode at low temperatures\cite{Chalker,Reimers2}.
Thus, the development of even  the spin SRO is largely
suppressed in the present model.

The results of the chirality-related quantities are given in Figs.5 and 6.
In Fig.5, we show the temperature and size dependence of
the mean local amplitude of the chiralities.
As can clearly be seen from the figure,
$\bar \chi$
has a  nonzero value even in the $T\rightarrow 0$ limit, while
$\bar \chi_{\rm tri}$ tends to zero.
This observation
indicates that the spins on  tetrahedron form locally non-coplanar
structures, while
the spins not belonging to  tetrahedron form locally coplanar structures.
In any case, the fact that the spins on tetrahedron form
the non-coplanar structures at low temperatures
sustaining the nontrivial chirality forms the basis of our following analysis.

Once establishing the existence of nontrivial local chirality,
the next obvious question is how these chiralities order with decreasing
temperature.
In Fig.6, we show
the size and temperature dependence of the chiral-glass
order parameter $q_\chi^{(2)}$. As is evident from the figure,
$q_\chi^{(2)}$ does not grow with decreasing temperature,
rapidly decreasing with increasing $L$. This indicates that
the chirality remains fluctuating until low temperature without
a finite-temperature transition.
Thus, in the case of $J_2=0$, although the chirality
certainly becomes  nontrivial locally, it remains fluctuating until
low temperatures  without exhibiting thermodynamic phase transition.
The absence of chiral transition in the $J_2=0$ case is easy to understand,
if one notes the fact that the {\it next-nearest-neighbor\/} interaction is 
necessary in order
to directly couple the neighboring tetrahedra on the kagom\'e layers 
at which
the chirality becomes nontrivial.

\subsection{The case of antiferromagnetic next-nearest-neighbor
interaction: $J_2/J_1=0.5$}

In this subsection, we consider the case in which the nnn
interaction on the kagom\'e layers is antiferromagnetic, fixing its
magnitude to be $J_2=0.5J_1$.
The inter-plane interaction $J'$ is set equal to $J_1$ for
the time being. The other choice of $J'$ will be considered
later in this subsection.

In Fig.7, we show the temperature and size dependence of 
the specific heat per spin.
In contrast to the $J_2=0$ case,
the data show double peaks, the higher one
at $T=T_{p1}\simeq 0.29J_1$ and the lower one at $T=T_{p2}\simeq 0.09J_1$.
Such double-peak feature in the specific heat has not been
observed in the pure kagom\'e Heisenberg AF\cite{Chalker,Reimers2}.
The size dependence
of these specific-heat peaks reveals that the peak heights eventually saturate
with $L$, suggesting that
both peaks are non-divergent; either a non-divergent singularity with
$\alpha <0$, or a regular peak without any singularity.

In Fig.8, we show the temperature and size dependence of 
the linear and nonlinear magnetic susceptibilities per spin. 
While the nonlinear susceptibility exhibits no appreciable anomaly 
as in the case of $J_2=0$, the linear susceptibility exhibits 
a clear cusp-like anomaly at $T/J_1\simeq 0.085$
close to  the lower specific-heat peak,
which has not been seen in the $J_2=0$ case.

The origin of the higher specific-heat peak may be seen from Fig.9,
where we show the temperature and size dependence of
the ferrimagnetic magnetizations associated with the
$q=0$ and the $\sqrt 3\times \sqrt 3$ modes,  $m_{0}$ and 
$m_{\sqrt 3}$, respectively.
One can see from the figures that
the $q=0$ mode is dominant over the
$\sqrt 3\times \sqrt 3$ mode, the SRO
of which begins to grow around the higher specific-heat peak
temperature $T_{p1}$. This indicates that the higher specific-heat peak
is associated with the development of the $q=0$ SRO of Heisenberg spins.
Closer inspection  of Fig.9 reveals that,
around the lower specific-heat peak temperature $T_{p2}$,
$m_{0}$ tends to be suppressed with further lowering the temperature.
We shall return to this point later.

Since our model is the isotropic Heisenberg model in 2D,
one generally expects
that there is no spin LRO and even $m_{0}$ should vanish 
in the $L\rightarrow \infty $ limit at any finite
temperature.
In Fig.10, we show the the temperature and size dependence of
the spin Binder ratio associated with the $q=0$ mode, $g_0$.
With increasing $L$, the data monotonically decreases toward zero, 
suggesting that there is indeed no ordering in the spin sector at any finite
temperature.

What happened around $T_{p2}$ or the susceptibility-cusp
temperature may be seen from Figs.11 and 12 where we show the
chirality-related quantities.
The temperature and size dependence of
the mean local amplitude of the chiralities are shown in Fig.11.
The data indicate that, as in the
$J_2=0$ case, the spins belonging to tetrahedron
form locally non-coplanar
structures sustaining nontrivial chirality.
A very interesting observation comes out from Fig.12, where we show
the temperature and size dependences of  the ferrimagnetic chiral
order parameter $m_{\chi f}$ and of
the chiral-glass order parameter $q_\chi ^{(2)}$. One can see 
from the figures that 
$m_{\chi f}$ and $q_\chi ^{(2)}$
grow rather sharply around the temperature close to $T_{p2}$ or  the
susceptibility-cusp temperature. This suggests that the
lower specific-heat peak is somehow correlated with the onset of
the ferrimagnetic order of the chirality.

In Fig.13, we show the temperature and size dependences of the
Binder ratio of the ferrimagnetic chiral order parameter, $g_{\chi f}$.
For smaller sizes, the calculated $g_{\chi f}$ for various $L$
tend to cross at $T=T_{\rm c}\simeq 0.082J_1$
but for larger sizes they tend to merge at $T\leq T_c$,
signaling the
occurrence of a phase transition of the chirality.
The estimated transition temperature $T_{\rm c}/J_1=0.082(2)$ is in rough
agreement with the susceptibility-cusp temperature estimated above,
and is slightly below the lower specific-heat peak
temperature $T_{p2}$. Meanwhile, there is no appreciable anomaly
in the specific heat just at
$T=T_{\rm c}\simeq 0.082J_1$: See the inset of Fig.7. 
A merging behavior of the Binder
ratio, without accompanying
the discernible specific-heat anomaly just
at $T_c$ but only with a
non-divergent peak slightly above $T_c$,
suggests that the
observed chirality transition might essentially be of the Kosterlitz-Thouless
(KT)-type\cite{KT}.

In order to investigate the nature of this chiral transition
at $T=T_c$ in more detail, we show in Fig.14(a) the $L$-dependence of
the ordering susceptibility $N_sm_{\chi f}^2$ associated with
the ferrimagnetic chiral order
on a log-log plot for several temperatures.
As can be seen from the figure, while the data at higher temperatures
exhibit the characteristic behaviors of the disordered phase,
bending down toward some finite values,
those at $T\leq T_c\simeq 0.082J_1$
lie on straight lines, exhibiting the behavior expected for the
KT-like critical phase with algebraically-decaying  correlations.
The estimated slope of
the plots, which should be equal to $2-\eta$
with $\eta $ being the critical-point decay exponent,
is shown in Fig.14(b) as a function of temperature.
More precisely, we plot the quantities
\begin{equation}
2-\eta (T,L,L^{'})=
\frac{\ln (m_{\chi f}^2 (L)/m_{\chi f}^2 (L^{'}))}{\ln (L/L^{'})},
\end{equation}
calculated for various combinations of $L$ and $L'$.
As can be seen from Fig.14(b),
the estimated $\eta $ is found to
be around $1/4$ at $T=T_c\simeq 0.082J_1$, which gradually decreases
(or $2-\eta $ increases) with decreasing temperature.
This again indicates that the chiral transition at $T=T_c$ is
the KT-type transition.

Indeed, there is a good reason to expect such a KT-type transition for
the present chiral transition. As mentioned,
chirality is an Ising like quantity taking values
either positive or negative at each upward triangles in the kagom\'e layer.
Since
these upward triangles form the triangular lattice 
in themselves, there is a close
similarity between the chirality ordering of the present model and
the ordering of the 2D Ising model on the triangular lattice.
If the triangular Ising model possesses the antiferromagnetic nn
interaction and ferromagnetic nnn interaction, the model
is known to exhibit a  three-sublattice % ferrimagnetic
$\sqrt 3\times \sqrt 3$
ordering at a
finite temperature via the KT-type transition characterized by
the exponent $\eta =1/4$\cite{Landau,Takayama,Fujiki,Miyashita}.
Since the chirality ordering of the present model with $J_2>0$
is essentially the staggered one, as is evident from the observed growth of
the  $\sqrt 3\times \sqrt 3$ component in Fig.12(a),
it would be no surprise
that the chirality ordering here is essentially of the KT-type.

There are two possibilities
concerning the $\sqrt 3\times \sqrt 3$
ordering pattern in the KT phase:
One is the cosine phase  characterized by
the sublattice magnetizations of the type $(+,+,-)$ {\it etc.\/},
and the other is the sine phase (or the partial disordered phase) 
characterized by the sublattice magnetizations of the type
$(+,0,-)$ {\it etc\/}. In order to see
which pattern is actually realized in the present model, 
we show in Fig.15
the calculated two-dimensional distribution of $(m_\chi^R, m_\chi^I)$,
at a temperature $T/J_1=0.044$ well below $T_c/J_1\simeq 0.082$. 
Fig.15 shows that the system predominantly stays at the phase $\theta$
being equal to
$\frac{\pi}{3}(n+\frac{1}{2})$, indicating that the
sine or the partial disordered state of chirality is realized.
In other words, the chirality is ordered
into the  $(+,0,-)$ pattern, keeping
one of three sublattices totally disordered.
The sixfold symmetry of the data shown in Fig.15 is a manifestation of the
sixfold degeneracy of the ordered state.
In fact, the antiferromagnetic ordering pattern of the Ising-like variables
on three triangular sublattices can be mapped onto
the ordering pattern of the 2D
ferromagnetic six-clock model\cite{Alexander}, which
is also known to exhibit
the KT transition at a finite temperature 
with the exponent $\eta =1/4$\cite{Jose}.

It should be noticed that the
triangular Ising AF with the ferromagnetic
nnn interaction 
exhibits another phase transition with decreasing temperature,
into the low-temperature phase with a finite
LRO.\cite{Landau,Takayama,Fujiki,Miyashita}
The exponent $\eta $
at this second phase transition point is believed to be
$1/9$\cite{Jose,Miyashita}. Therefore,
we search for this second phase transition into the
long-range-ordered state in our present model.
 From Fig.14(b), one sees that, 
even in the low-temperature regime where the estimated $\eta$
comes down to $1/9$,
there is no sign that the system exhibits the second
transition into the long-range-ordered state
where $\eta$ should vanish. 
In other quantities such as the specific heat or the susceptibility,
we do not find
any evidence of the second phase transition. 
Presumably, severe frustration inherent to the present model might
hinder the
onset of the true chiral LRO. However, since our low-temperature data
are limited due to the difficulty of thermalization, we 
cannot completely exclude the possible occurrence of such a second
phase transition in our model. 

In fact, the value of $m_{\chi f}$, 
extrapolated to $T=0$ in Fig.12(a) and normalized by
$\bar \chi (T\rightarrow 0)$, is only
12\% of the value
expected if the chirality is fully ordered into the sine  pattern. 
Likewise, the value of $q_\chi^{(2)}$, 
extrapolated to $T=0$ in Fig.12(b) and normalized by the appropriate 
powers of 
$\bar \chi (T\rightarrow 0)$, is only  20\% of the value
expected if the chirality is fully frozen over the system.
These observations indicate that, even in the ordered state, chiralities 
are still strongly
fluctuating, only a fraction of them taking part in the chirality ordering.

We also mention here that the decrease of the $q=0$ Fourier
magnetization
below $T_c$ observed in Fig.9(a) 
could be understood if one notes the fact  that the perfect $q=0$
spin order tends to compete with the local non-coplanarity of spins.
In fact, the antiferromagnetic
nn interaction $J_1$ prefers the planar $120^\circ$ spin
structure at each downward triangle, which induces the planar state
for the $q=0$ spin state.
Hence, the onset of the KT order of the chirality below $T_c$
necessarily compete with
the $q=0$ spin order, and suppresses it. 
Such suppression of the $q=0$ spin order below $T_c$
can also be seen in the behavior of the chiral Binder ratio in
Fig.10.

     So far, we have considered the case where the inter-plane interaction
$J'$ is equal to the intra-plane nn interaction $J_1$.
In real experimental systems, however, such equality is not expected
in general. Therefore,
in order to examine
the possible effect of varying the inter-plane coupling,
we deal with the case
$J'=0.5J_1$ (and $J_2=0.5J_1$ as before)
in the remaining part of this subsection.

We find that most of the calculated
physical quantities behave similarly to
those in the $J'=J_1$ case shown above. As an example,
we show in Figs.16-18 the temperature and size dependences of
the specific heat, the linear susceptibility, and the chiral Binder ratio.
These data indicate that the system exhibits a
chiral KT transition at $T_c/J_1=0.029(3)$. The exponent $\eta $
at $T=T_c$ is again consistent with the
KT value $\eta =1/4$ as can be seen from Fig.19.
We find no evidence of the occurrence of the
second transition into the
long-range-ordered phase. As is evident from Fig.20, the
KT order of chirality
is the sine or the partial disordered state.
All these features are qualitatively the same as those observed in the
$J'=J_1$ case. Hence, we conclude that the condition $J'=J_1$
is irrelevant,
and the KT-type chiral transition identified
in the $J'=J_1$ case would occur generically for other $J'$ values
so long as
the nnn coupling $J_2$ is antiferromagnetic.

\subsection{The case of ferromagnetic next-nearest-neighbor interaction: 
$J_2/J_1=-0.5$}

In this subsection,
we consider the case in which the nnn
interactions in the
kagom\'e layers is {\it ferromagnetic\/}, fixing its
magnitude to be $J_2=-0.5J_1$.
The antiferromagnetic
inter-plane interaction $J'$ is again set equal to $J_1$,
{\it i.e.\/}, we assume $J'=J_1$ throughout this subsection.

In Fig.21, we show
the temperature and size dependence of the specific heat per spin.
The data show double peaks at $T/J_1\simeq 0.52$ and at $T/J_1\simeq 0.02$.
In Fig.22, we show
the temperature and size dependence of the linear and nonlinear
susceptibilities per spin. Both the linear and 
nonlinear susceptibilities
exhibit no appreciable anomaly in the investigated temperature
range. This should be contrasted to the $J_2>0$ case
where the linear susceptibility exhibits a clear cusp-like anomaly
as shown in Fig.8(a).

In order to get information about the spin SRO,
we show in Fig.23 the temperature and size dependence of
the ferrimagnetic magnetizations associated with the
$q=0$ and the $\sqrt 3\times \sqrt 3$ modes, $m_{0}$ and 
$m_{\sqrt 3}$, respectively. One can immediately see from
these figures that, in contrast to the $J_2>0$ case shown in Fig.9,
the $\sqrt 3\times \sqrt 3$ mode is stabilized over the $q=0$ mode,
reflecting the change in the sign of $J_2$. The higher specific-heat
peak is well correlated with the onset of the $\sqrt 3\times \sqrt 3$
spin SRO.

The chirality-related quantities are shown in Figs.24 and 25.
Fig.24 indicates that, as in  other cases studied, 
the spins belonging to tetrahedron form locally non-coplanar
structures, sustaining nontrivial chirality.
Although $q_{\chi f}^{(2)}$ for a fixed size
grows with decreasing temperature below $T/J_1\simeq 0.1$ 
as can be seen from Fig.25, probably reflecting the growth of 
$\bar \chi$ there,
this tendency is more and more suppressed with increasing $L$. 
Furthermore,
$q_{\chi f}^{(2)}$ values themselves remain small
compared with those for $J_2>0$, and decreases rapidly with
increasing $L$ in contrast to the $J_2>0$ case.
Hence, in the $J_2<0$ case, 
we conclude that there is no thermodynamic phase transition
of the chirality, even though the specific heat has double peaks
shown in Fig.21.
The orgin of the lower specific-heat peak is not clear at the moment.

\section{Discussion and Summary}

We have studied by means of MC simulations the ordering
properties of
the antiferromagnetic Heisenberg model on a pyrochlore slab.
Due to the tetrahedron-based structure of this lattice,
thermodynamic properties of
the antiferromagnetic Heisenberg model on a pyrochlore slab
are quite different from those of the well-studied antiferromagnetic
Heisenberg model on the kagom\'e lattice.
In the case of the kagom\'e Heisenberg AF, the spin configuration
selected at low temperatures is a coplanar one, while in the case of
the pyrochlore-slab AF, the spin configuration
selected at low temperatures is a non-coplanar one sustaining the
nontrivial chirality. We have performed a detailed numerical study
of the chiral properties of the model.
Among others, we have found that,
when the nnn interaction on the kagom\'e layers
is antiferromagnetic, the chiralities exhibit a
KT-type phase transition at a finite temperature,
with keeping the Heisenberg spins being paramagnetic.
The KT ordered state of chirality
is characterized by the
sine-type $\sqrt 3\times \sqrt 3$ order (or the partial disorder).
To the authors' knowledge, our present finding is the first case of
{\it chiral\/} KT transition without accompanying any spin order.

Our present model is expected to capture some
essential geometrical ingredients of SCGO, and indeed seems to account for
some of the experimental features, {\it e.g.\/},
the occurrence of a finite-temperature transition 
above all, 
the cusp-like anomaly observed
in the linear susceptibility\cite{Ramirez},
or the existence of a  specific-heat peak
slightly above the transition temperature\cite{Ramirez}.
However, one immediately sees that some other experimental features
remain unexplained.
Experimentally, the negative divergence of the nonlinear susceptibility
is observed at $T_c$\cite{Ramirez}, which was not observed in our model.

One common property characterizing the SG-like freezing
transition, including the one observed in SCGO, 
is the magnetic irreversibility, often detected as a notable difference
between the field-cooled (FC) and zero-field-cooled (ZFC) dc
susceptibilities\cite{Ramirez}.
In order to detect the possible difference between the FC and ZFC
susceptibilities in the present model, we have also measured these
quantities by MC simulation in the case of $J_2=0.5J_1$ and $J'=J_1$
where our model is found to exhibit 
a chiral KT transition at $T=T_c\simeq 0.082J_1$.
The system is gradually cooled or warmed in an applied field according to
the standard heat-bath updating without the temperature-exchange
process. Note that here the system is not necessarily
equilibrated fully at each temperature.
The result shown in Fig.26, however, reveals that
there is no appreciable difference between the FC and ZFC susceptibilities
even below $T_c\simeq 0.082J_1$.
Therefore, our present model still fails to reproduce an important
spin-glass feature observed in real SCGO.

Thus, while our present model might well capture some aspects
of the ordering of SCGO so far neglected or unappreciated,
it is still inadequate
in fully describing the experimental result. Some
important aspects which have not been taken into account
in our present model,
{\it e.g.\/}, the existence of quenched randomness, quantum fluctuations,
or weak magnetic anisotropy {\it etc.\/}, 
might play an essential role in real SCGO.
Further MC simulations taking account of some of these effects 
are now in progress.

The numerical calculation was performed on the Hitachi SR8000 at the
supercomputer center, ISSP, University of Tokyo.

% References

\end{document}

\sloppy
\begin{figure}
\begin{center}
\includegraphics[scale=0.6]{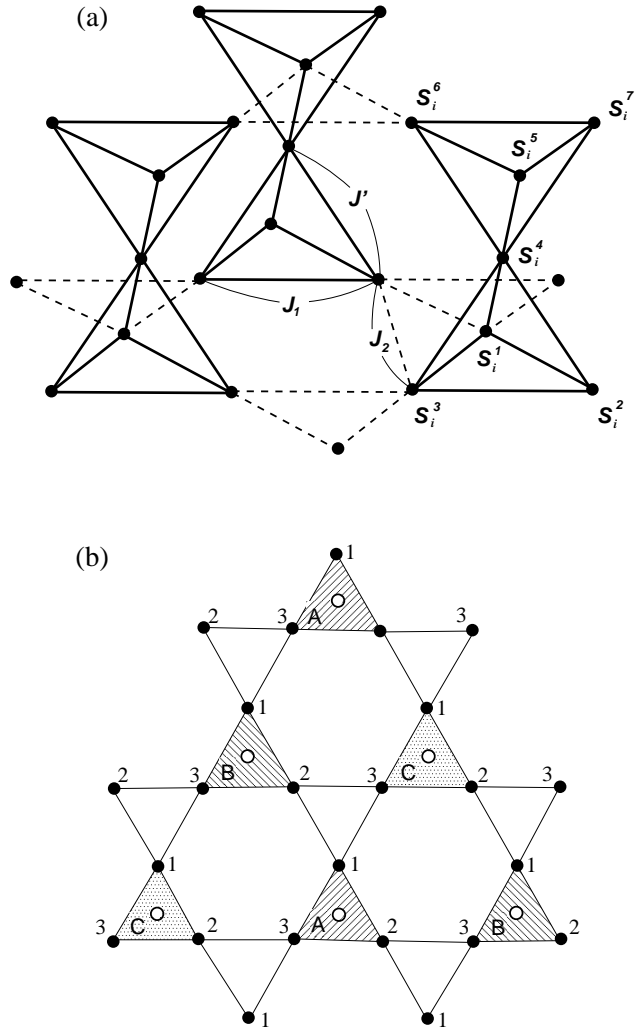}
\caption{A pyrochlore-slab lattice: 
(a) The unit cell of the lattice consists of
two corner-sharing tetrahedra  illustrated  by the solid lines.
Each unit cell
contains seven sites numbered from 1 to 7.
Within the kagom\'e layers, there are the nearest-neighbor
antiferromagnetic interaction $J_1$ and the next-nearest-neighbor
interaction $J_2$, while, between the kagom\'e layers and the
triangular layer sandwiched by the two kagom\'e layers,
there is the antiferromagnetic interaction
$J'$ along the edge of tetrahedron: (b) The lower kagom\'e
layer consisting of the sites 1, 2 and 3, denoted by the
solid circle. The open circle denotes the apical site of
tetrahedron (site 4).
The upward triangle
corresponds to the bottom plane of tetrahedron, which form
the triangular lattice.
These upward triangles are grouped
into three types denoted by $A$, $B$ and $C$, each forming the three
inter-penetrating triangular sublattices.}
\end{center}
\end{figure}

\newpage

\begin{figure}
\begin{center}
\includegraphics[scale=0.8]{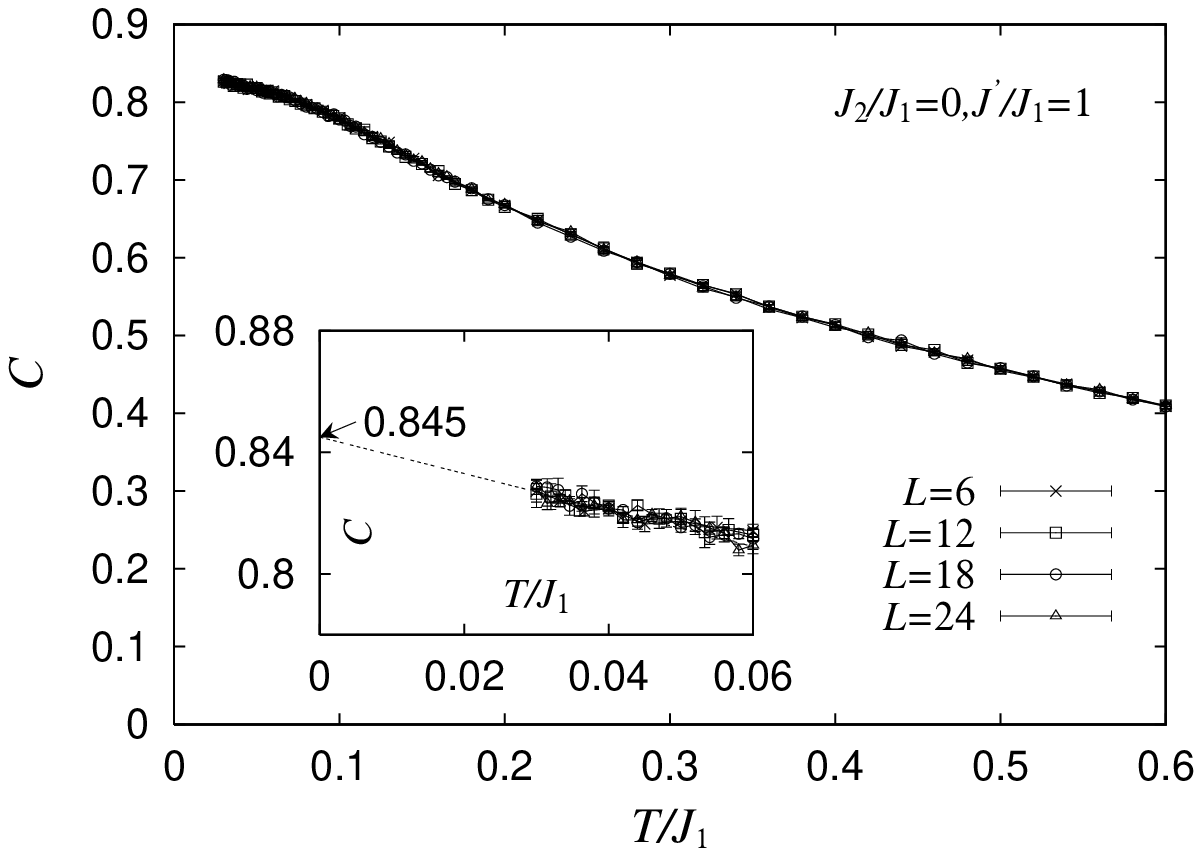}
\caption{The temperature and size dependence of the specific heat
per spin for the case $J_2/J_1=0$ and $J'/J_1=1$. The inset is a
magnified view of the low-temperature regime.}
\end{center}
\end{figure}

\newpage

\begin{figure}
\begin{center}
\includegraphics[scale=0.8]{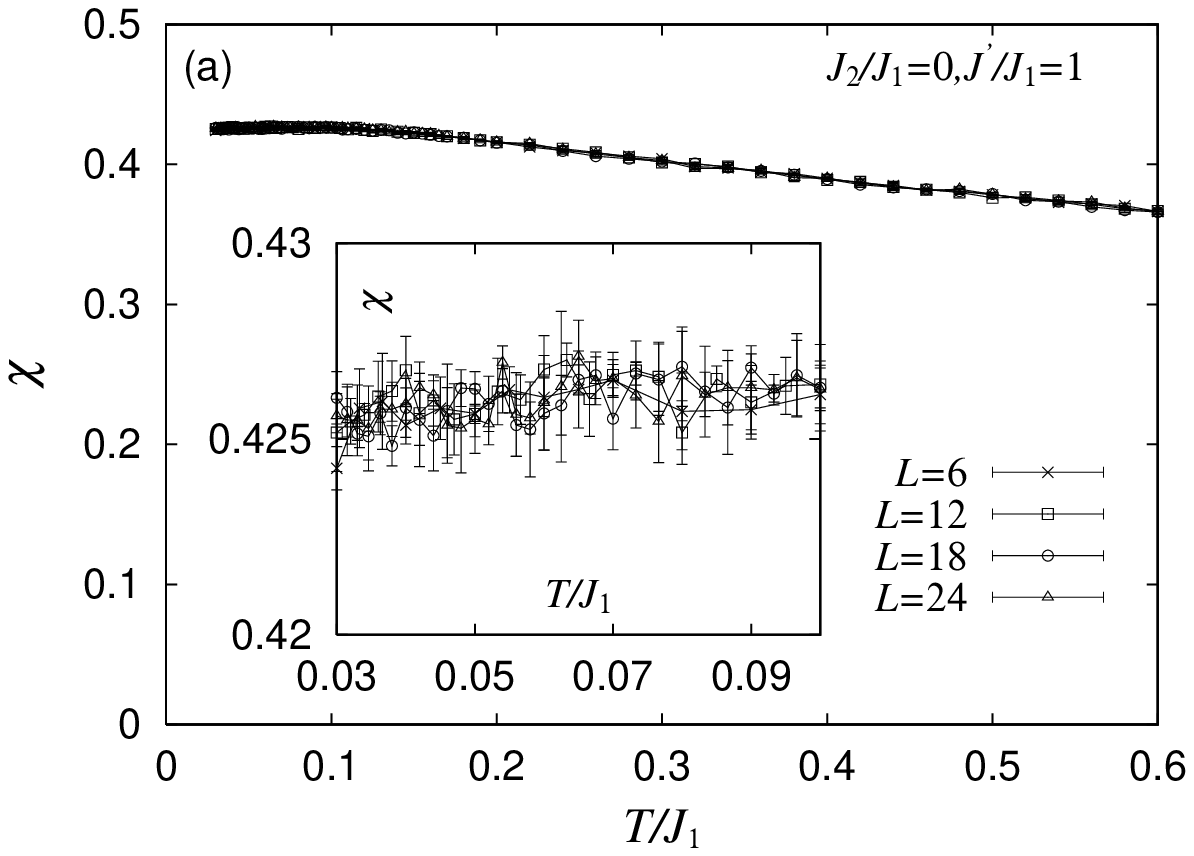}
\hspace*{1mm}
\includegraphics[scale=0.79]{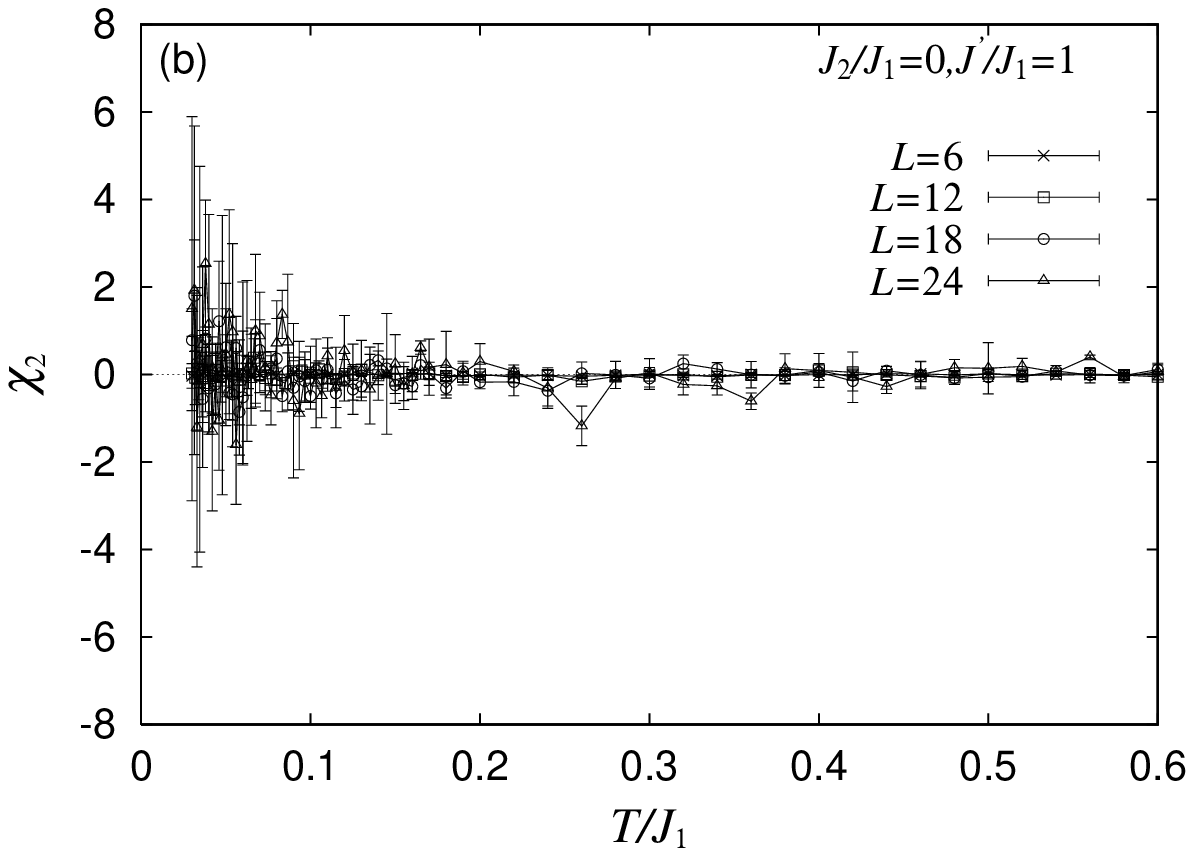}
\caption{The temperature and size dependence of (a) the linear 
susceptibility per spin, and (b) the nonlinear susceptibility
per spin,  for the case $J_2/J_1=0$ and $J'/J_1=1$.
The inset is a
magnified view of the low-temperature regime.}
\end{center}
\end{figure}

\newpage

\begin{figure}
\begin{center}
\includegraphics[scale=0.8]{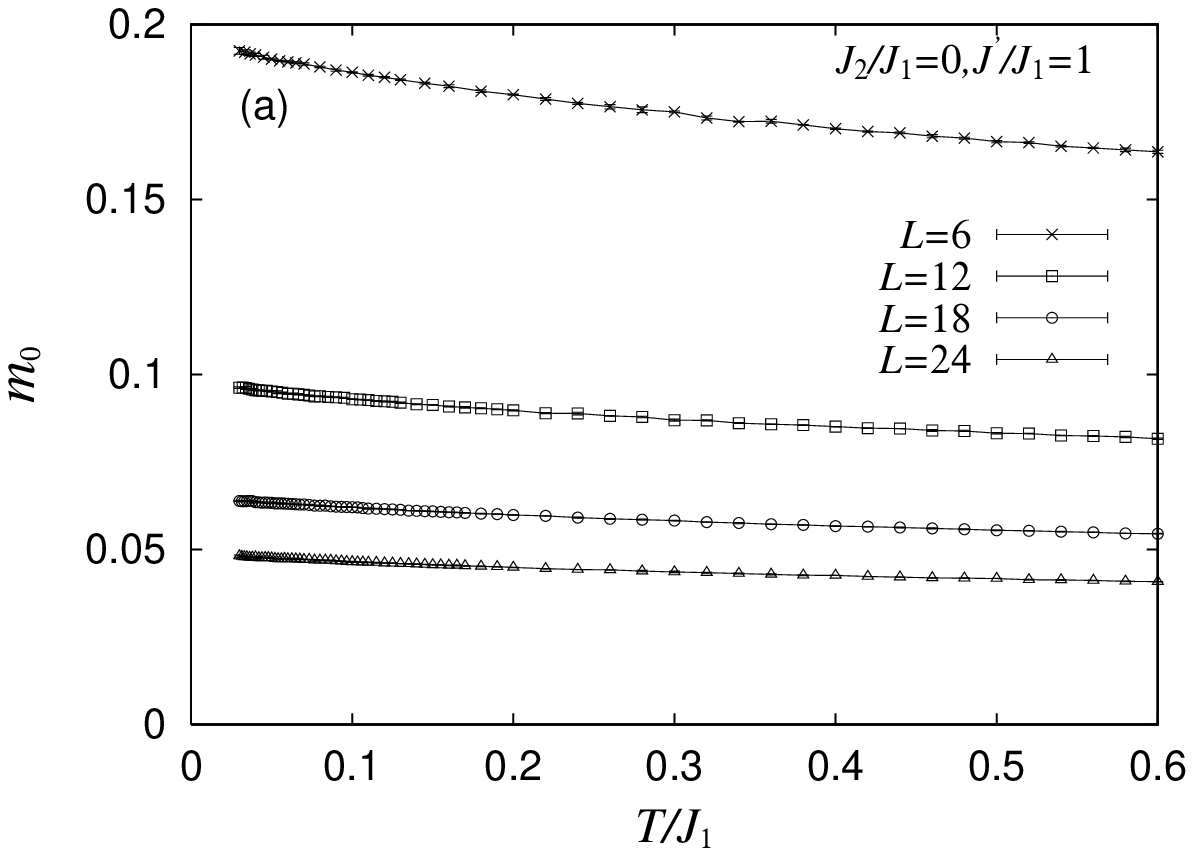}
\includegraphics[scale=0.8]{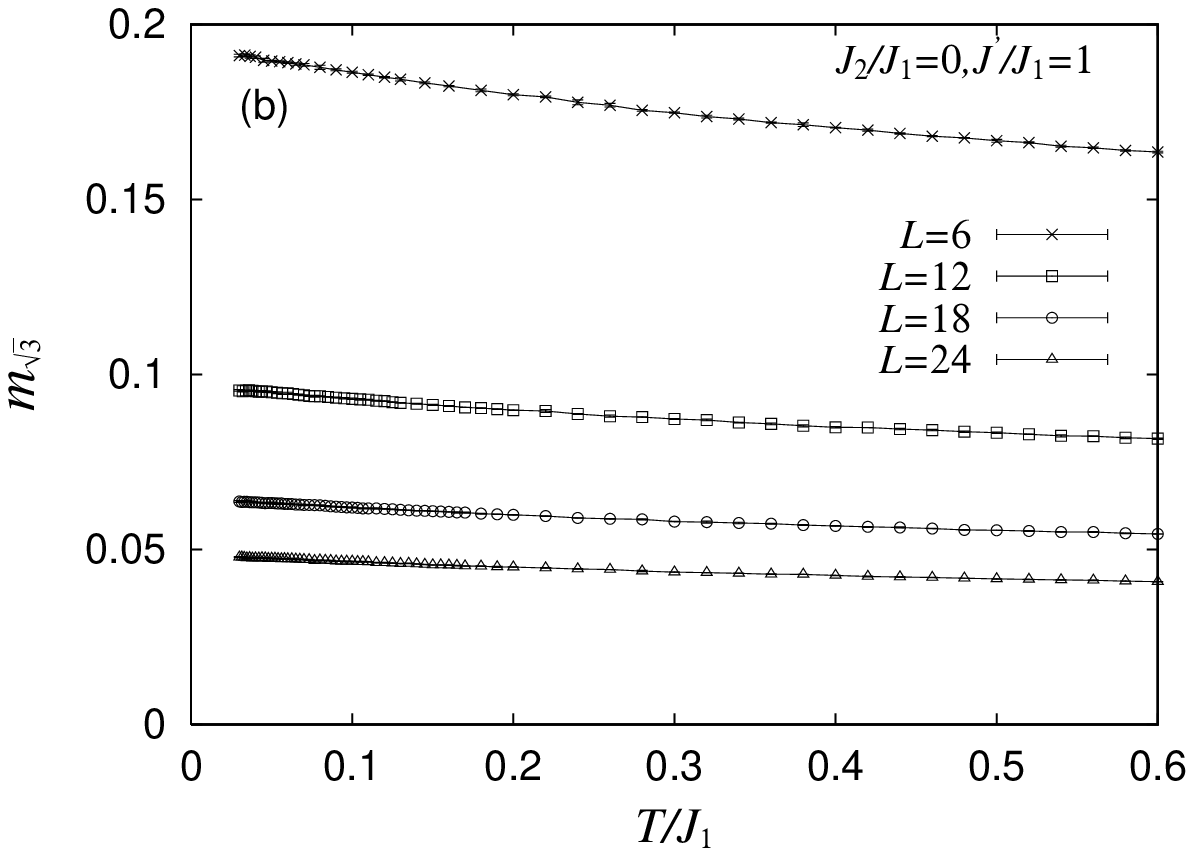}
\caption{The temperature and size dependence of the
ferrimagnetic magnetization per spin
associated with (a) the $q=0$ mode, and (b) the $\sqrt 3\times \sqrt 3$
mode, for the case $J_2/J_1=0$ and $J'/J_1=1$.}
\end{center}
\end{figure}

\newpage

\begin{figure}
\begin{center}
\includegraphics[scale=0.8]{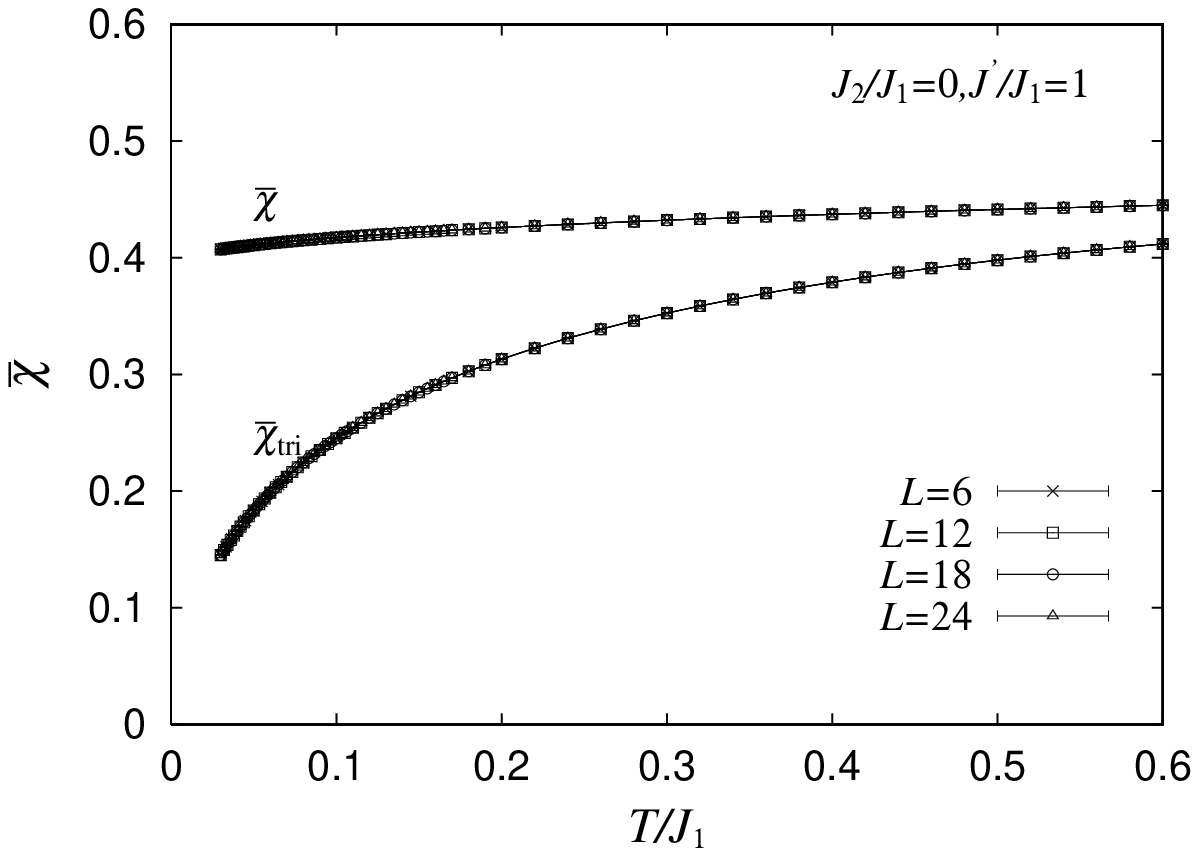}
\caption{The temperature and size dependence of the mean local
amplitude of the chirality
for the case $J_2/J_1=0$ and $J'/J_1=1$. $\bar \chi$ and
$\bar \chi_{{\rm tri}}$ represent the local chirality defined at the
upward and downward triangles on the kagom\'e layer, respectively.}
\end{center}
\end{figure}

\begin{figure}
\begin{center}
\includegraphics[scale=0.8]{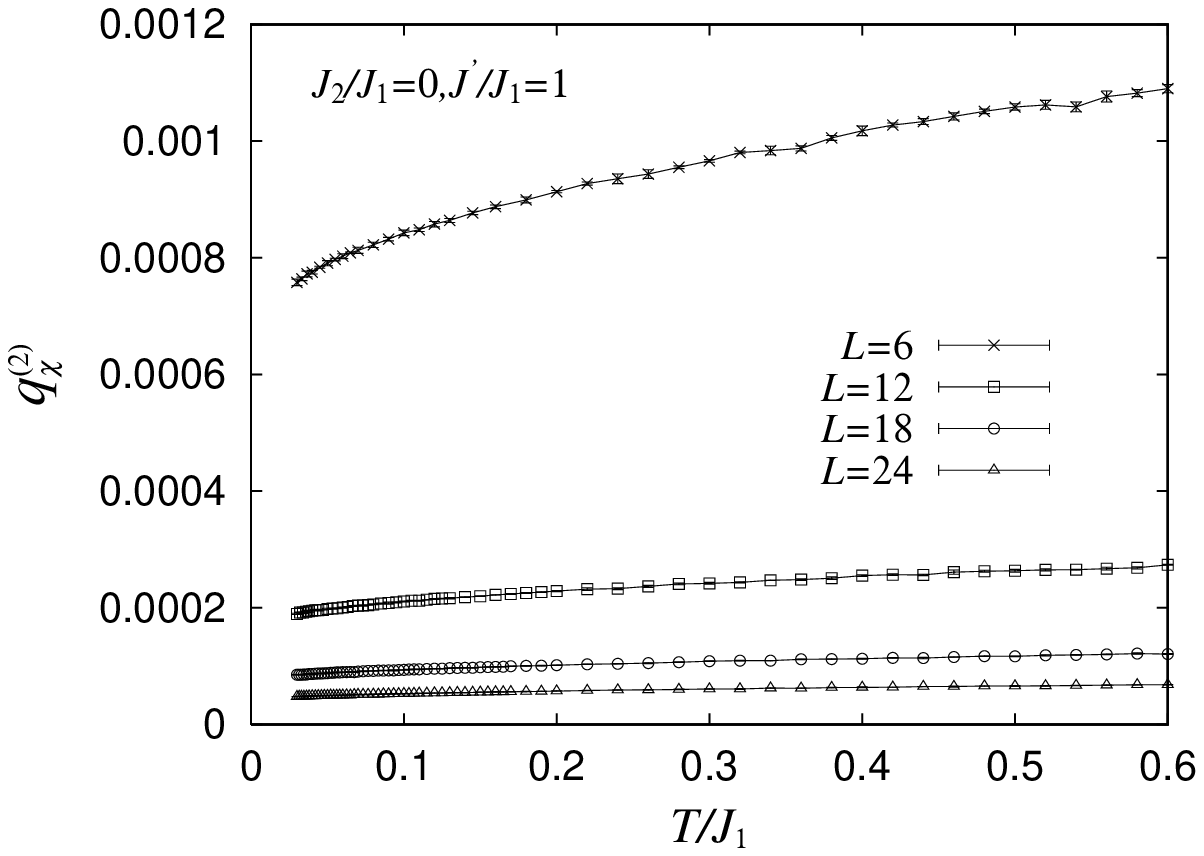}
\caption{The temperature and size dependence of the chiral-glass
order parameter
for the case $J_2/J_1=0$ and $J'/J_1=1$.}
\end{center}
\end{figure}

\newpage

\begin{figure}
\begin{center}
\includegraphics[scale=0.8]{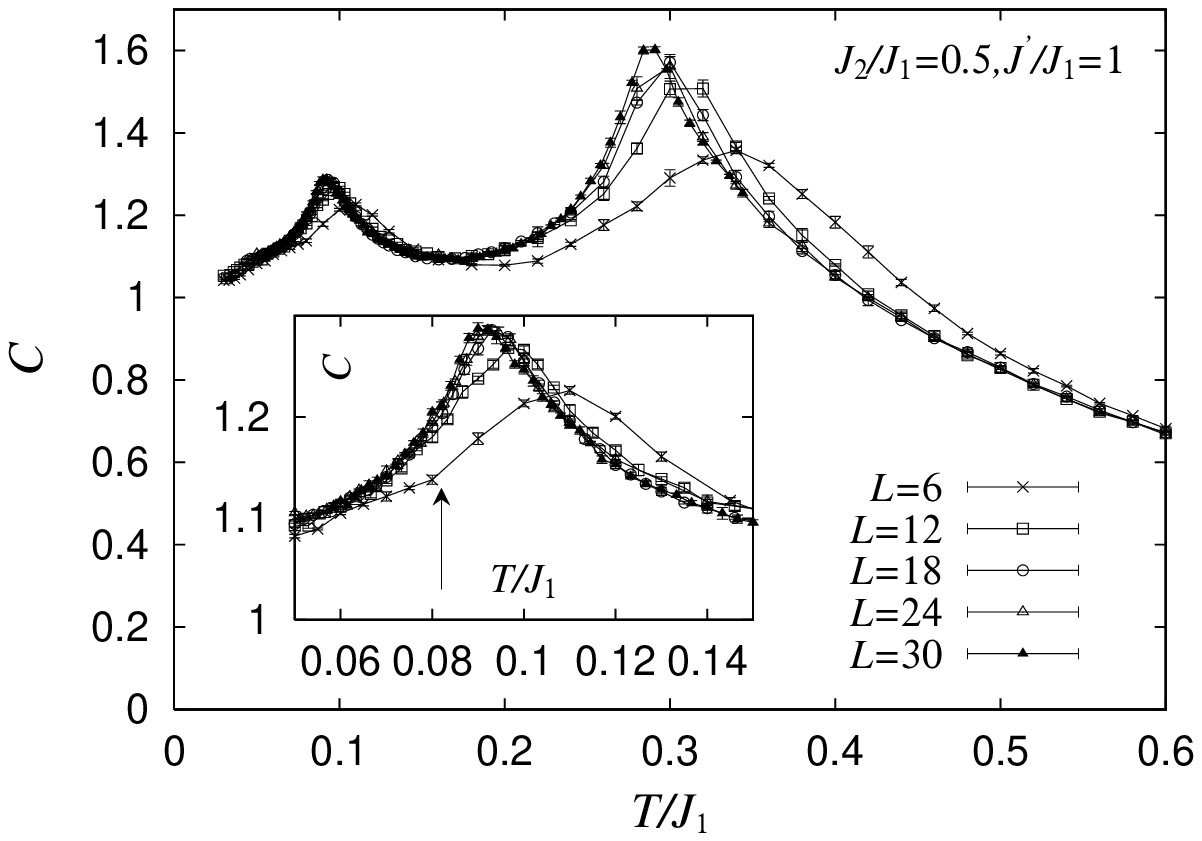}
\caption{The temperature and size dependence of the specific heat
per spin for the case $J_2/J_1=0.5$ and $J'/J_1=1$. The inset is a
magnified view of the low-temperature regime. The arrow indicates
the transition point.}
\end{center}
\end{figure}

\newpage

\begin{figure}
\begin{center}
\includegraphics[scale=0.8]{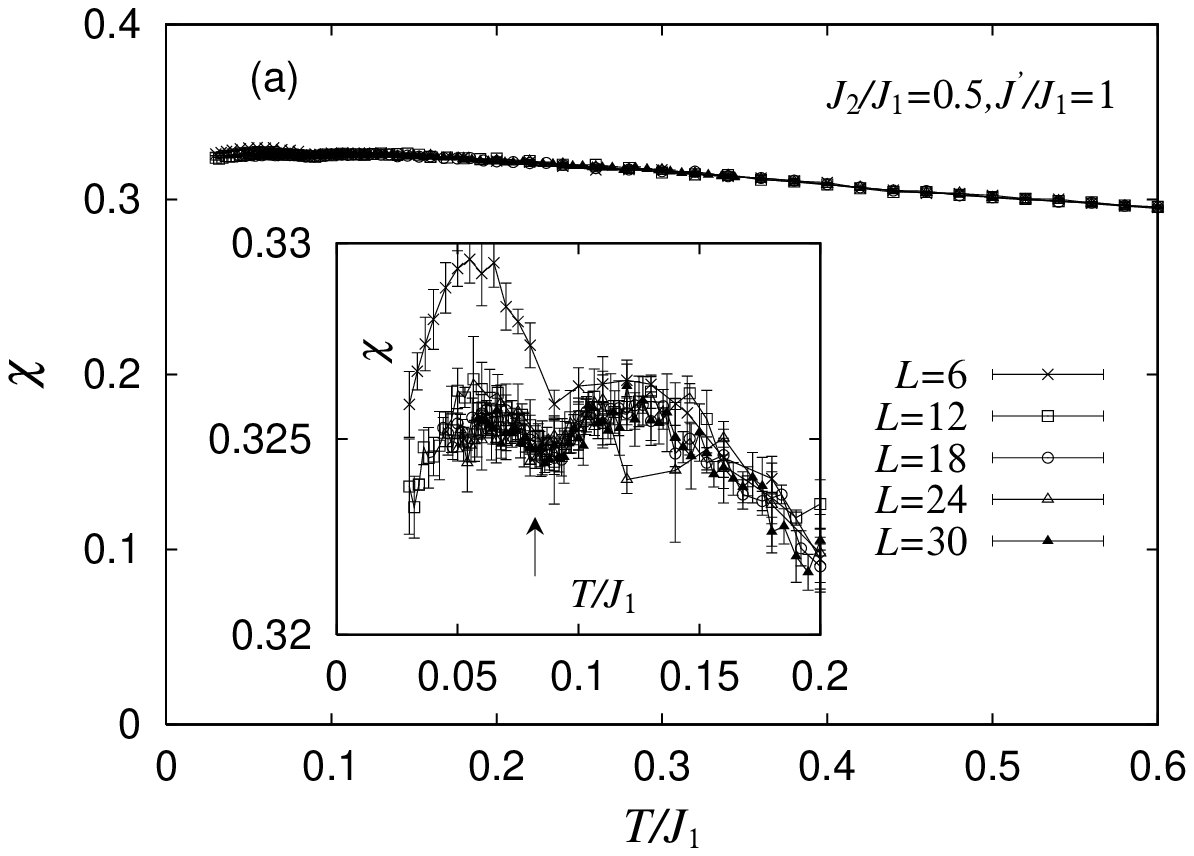}
\hspace*{1mm}
\includegraphics[scale=0.79]{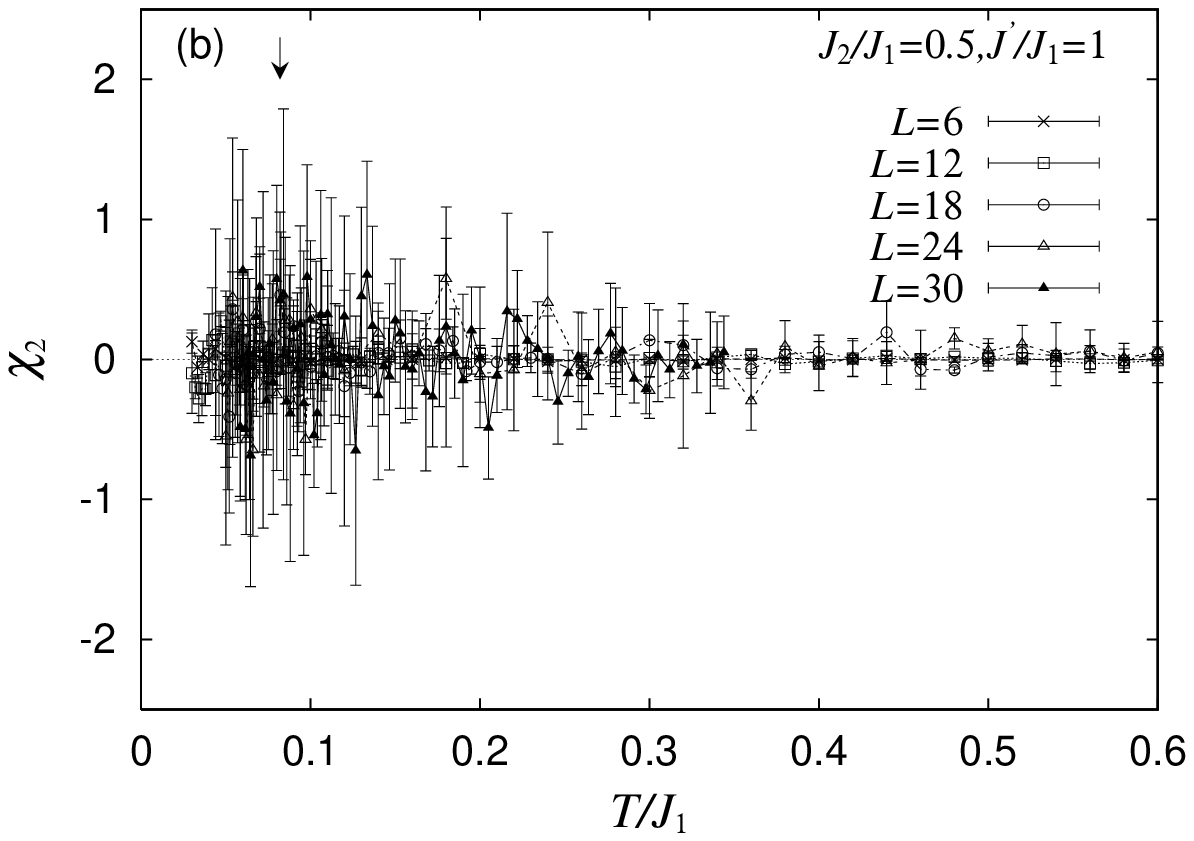}
\caption{The temperature and size dependence of the (a) linear
susceptibility per spin, 
and (b) the nonlinear susceptibility per spin, 
for the case $J_2/J_1=0.5$ and $J'/J_1=1$.
The inset is a
magnified view of the low-temperature regime.
The arrow indicates the transition point.}
\end{center}
\end{figure}

\newpage

\begin{figure}
\begin{center}
\includegraphics[scale=0.8]{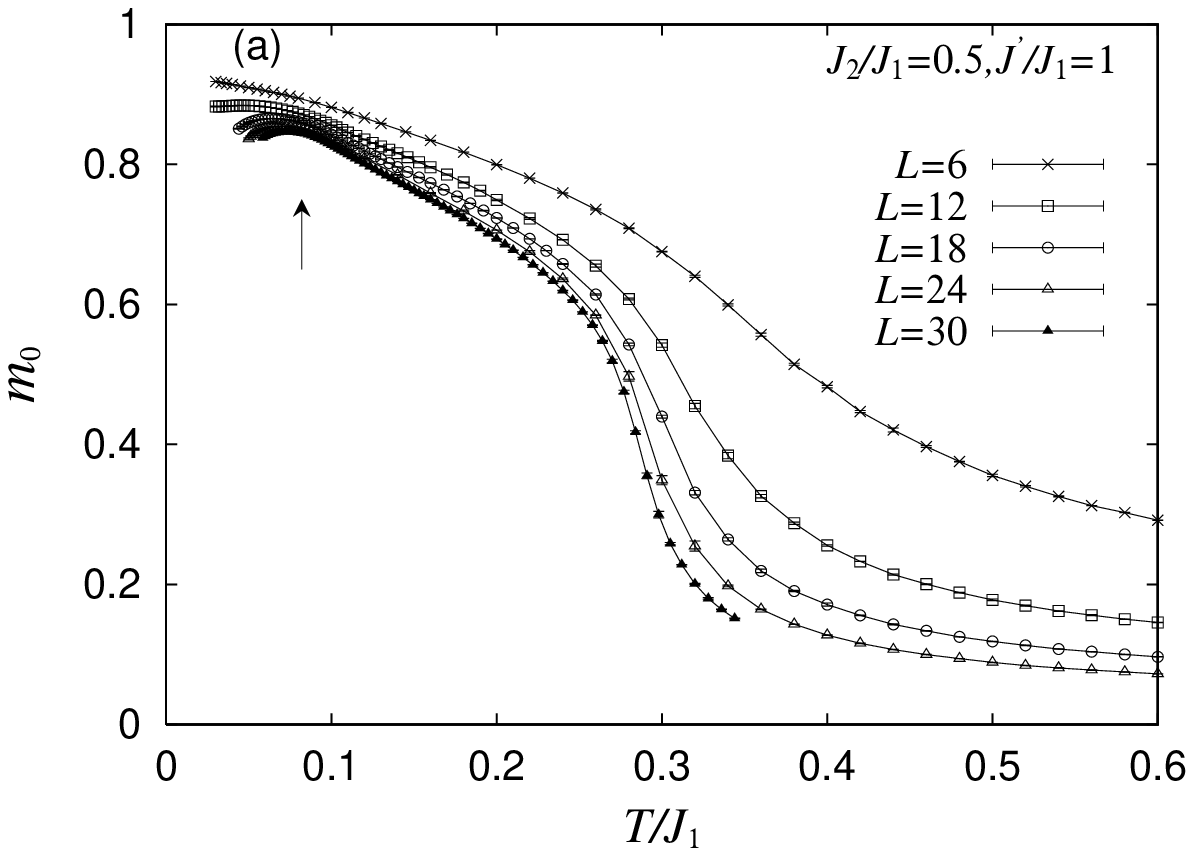}
\includegraphics[scale=0.8]{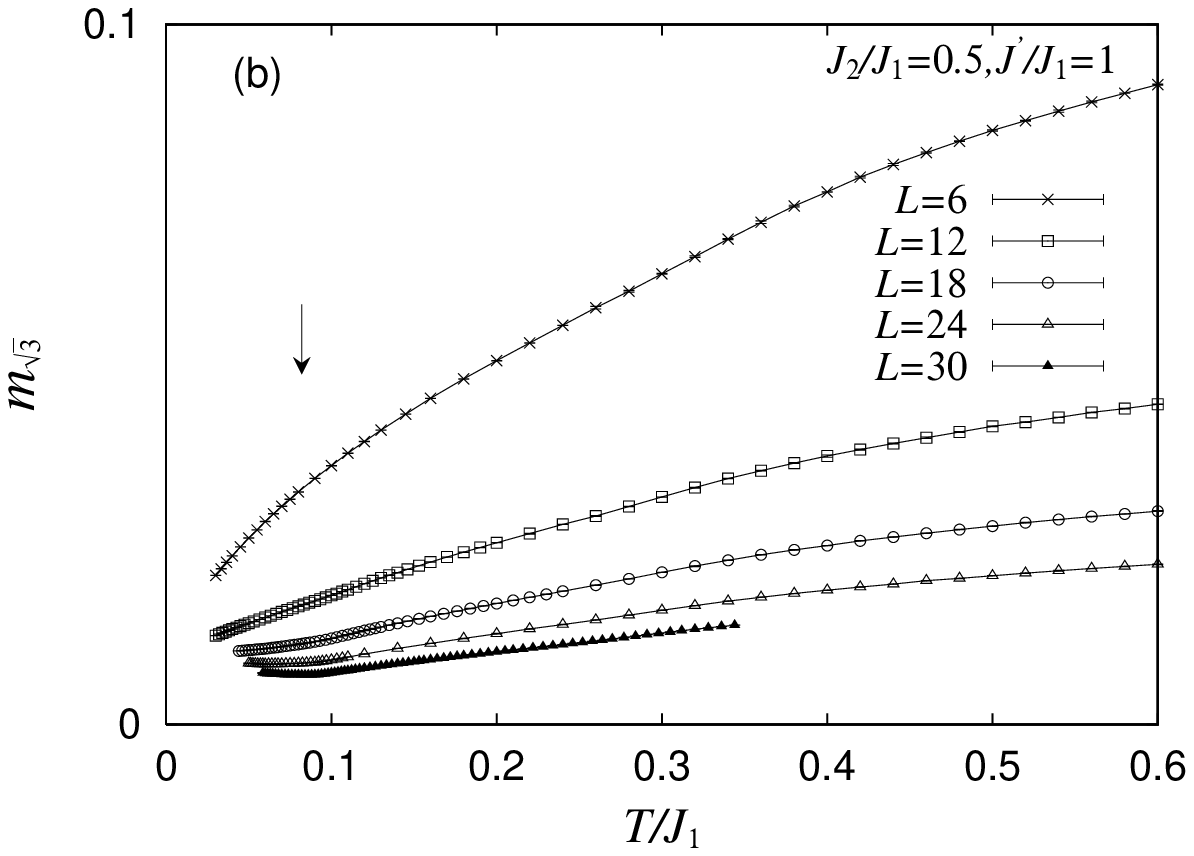}
\caption{The temperature and size dependence of the
ferrimagnetic magnetization per spin
associated with  (a) the $q=0$ mode, and (b) the $\sqrt 3\times \sqrt 3$
mode, for the case $J_2/J_1=0.5$ and $J'/J_1=1$.
The arrow indicates the transition point.}
\end{center}
\end{figure}

\begin{figure}
\begin{center}
\includegraphics[scale=0.8]{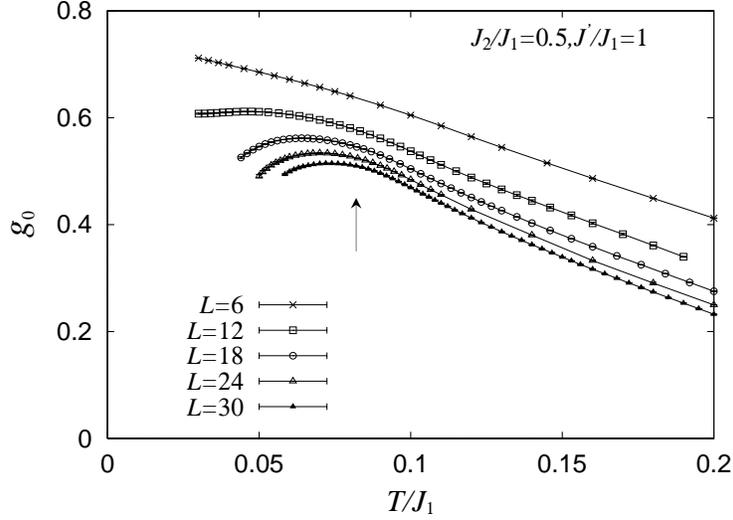}
\caption{The temperature and size dependence of the spin
Binder ratio associated with the $q=0$ mode
for the case $J_2/J_1=0.5$ and $J'/J_1=1$.
The arrow indicates the transition point.}
\end{center}
\end{figure}

\begin{figure}
\begin{center}
\includegraphics[scale=0.8]{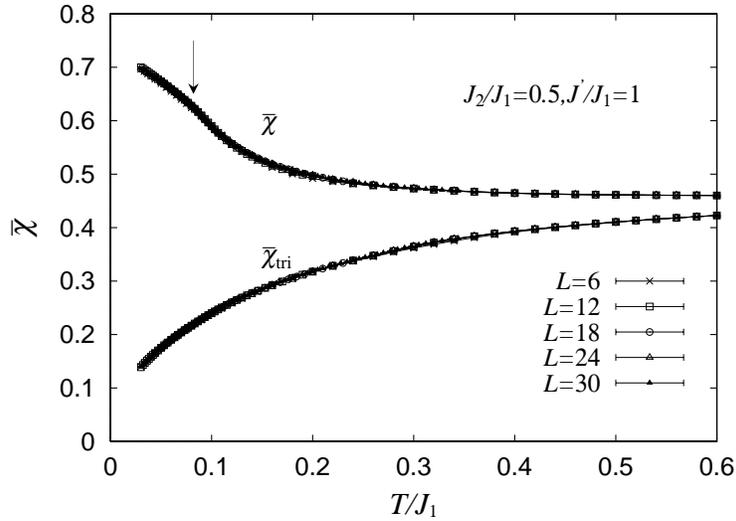}
\caption{The temperature and size dependence of the mean local
amplitude of the chirality
for the case $J_2/J_1=0.5$ and $J'/J_1=1$. $\bar \chi$ and
$\bar \chi_{{\rm tri}}$ represent the local chirality defined at the
upward and downward triangles on the lower kagom\'e layer.
The arrow indicates the transition point.}
\end{center}
\end{figure}

\newpage

\begin{figure}
\begin{center}
\includegraphics[scale=0.8]{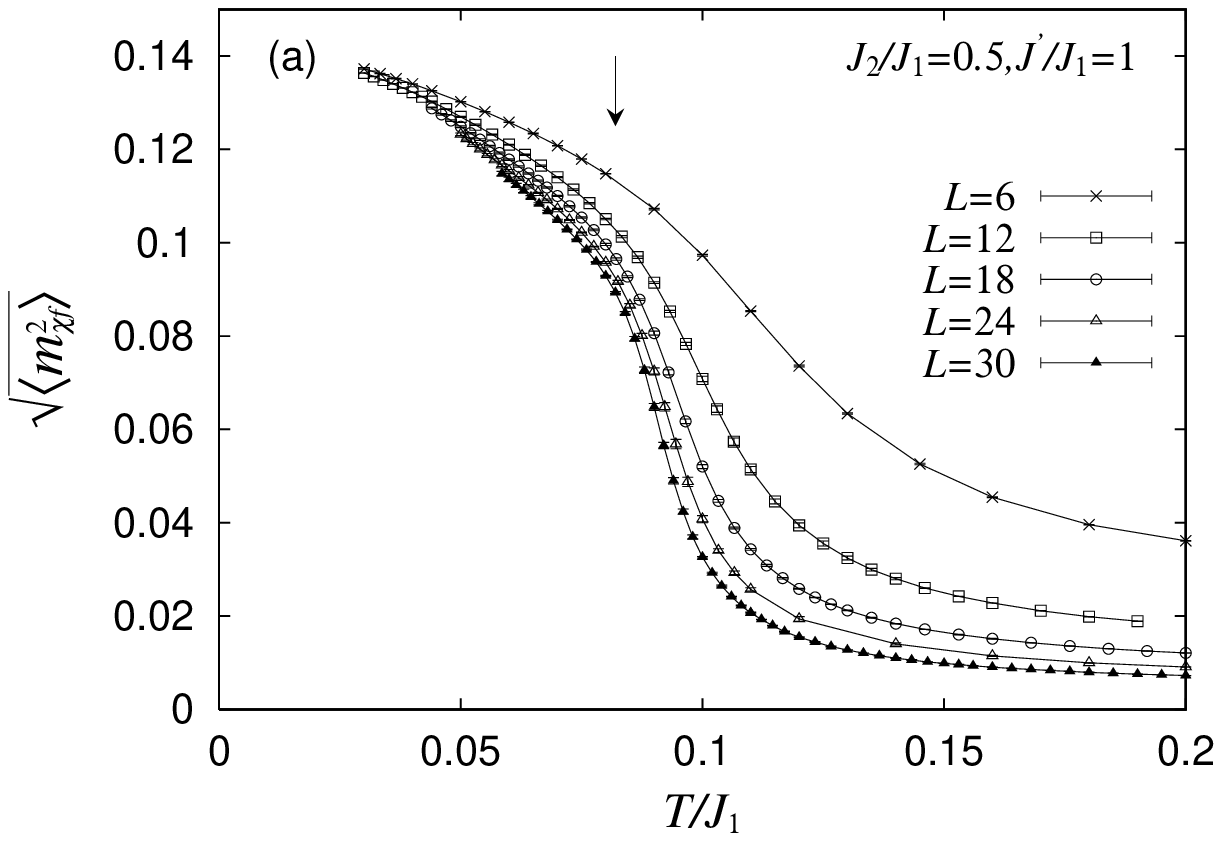}
\includegraphics[scale=0.8]{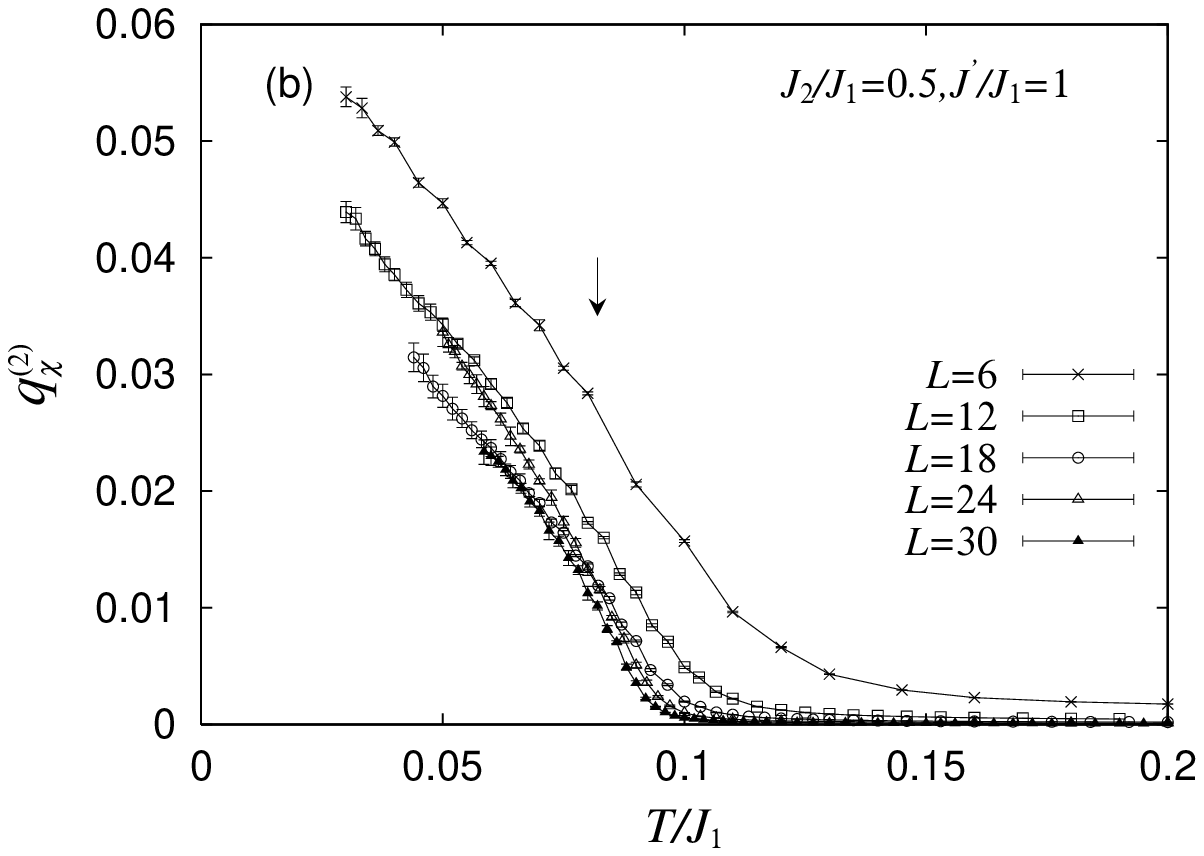}
\caption{The temperature and size dependence of (a) the
ferrimagnetic (staggered) chirality, and (b) the
chiral-glass order parameter,
for the case $J_2/J_1=0.5$ and $J'/J_1=1$.
The arrow indicates the transition point.}
\end{center}
\end{figure}

\newpage

\begin{figure}
\begin{center}
\includegraphics[scale=0.8]{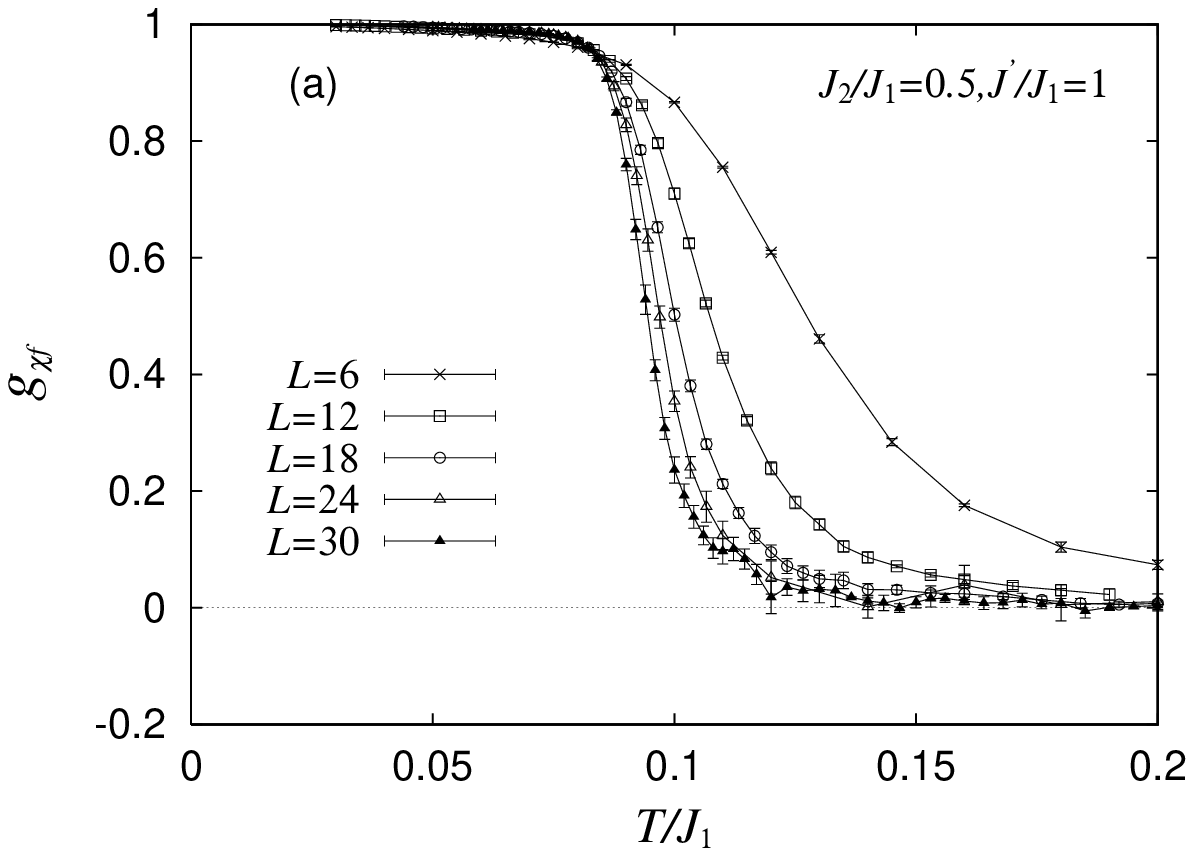}
\includegraphics[scale=0.8]{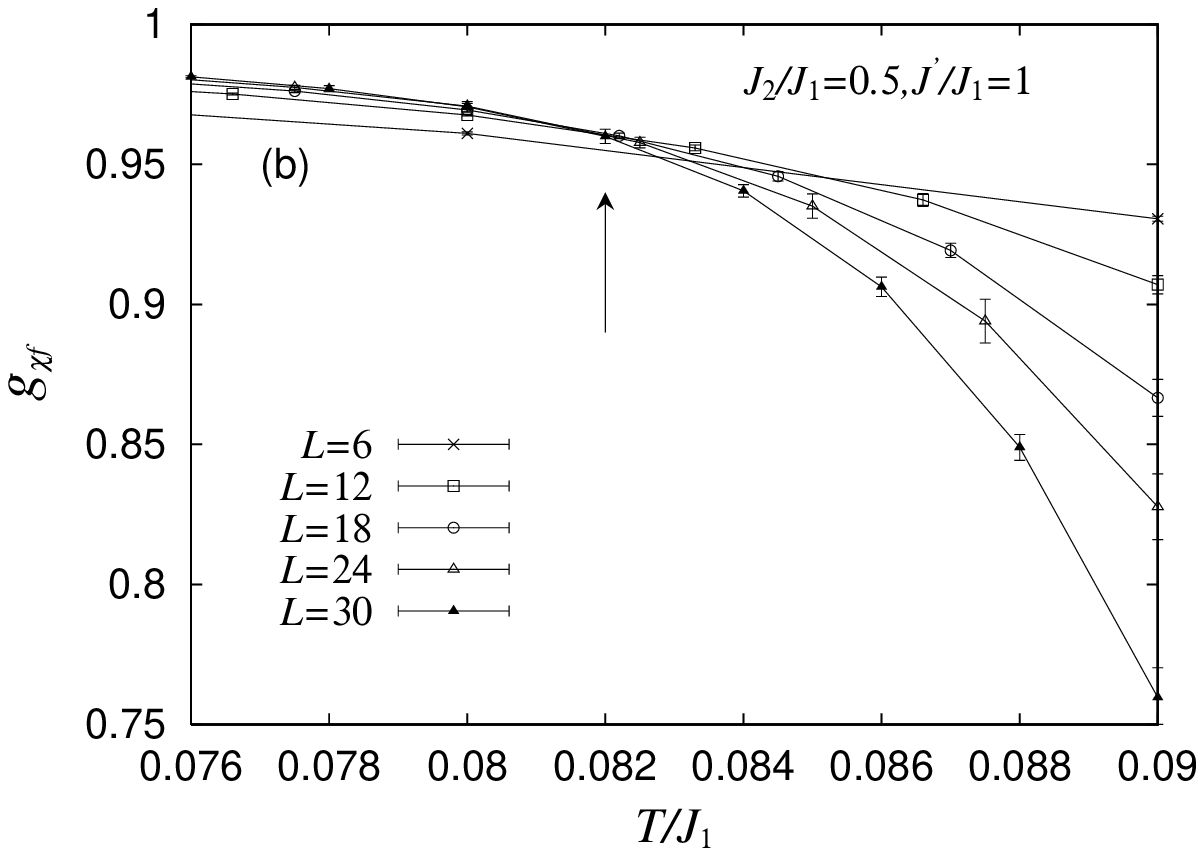}
\caption{(a) The temperature and size dependence of the chiral
Binder ratio for the case $J_2/J_1=0.5$ and $J'/J_1=1$.
(b) A magnified view of the transition region.
The arrow indicates
the transition point.}
\end{center}
\end{figure}

\newpage

\begin{figure}
\begin{center}
\includegraphics[scale=0.8]{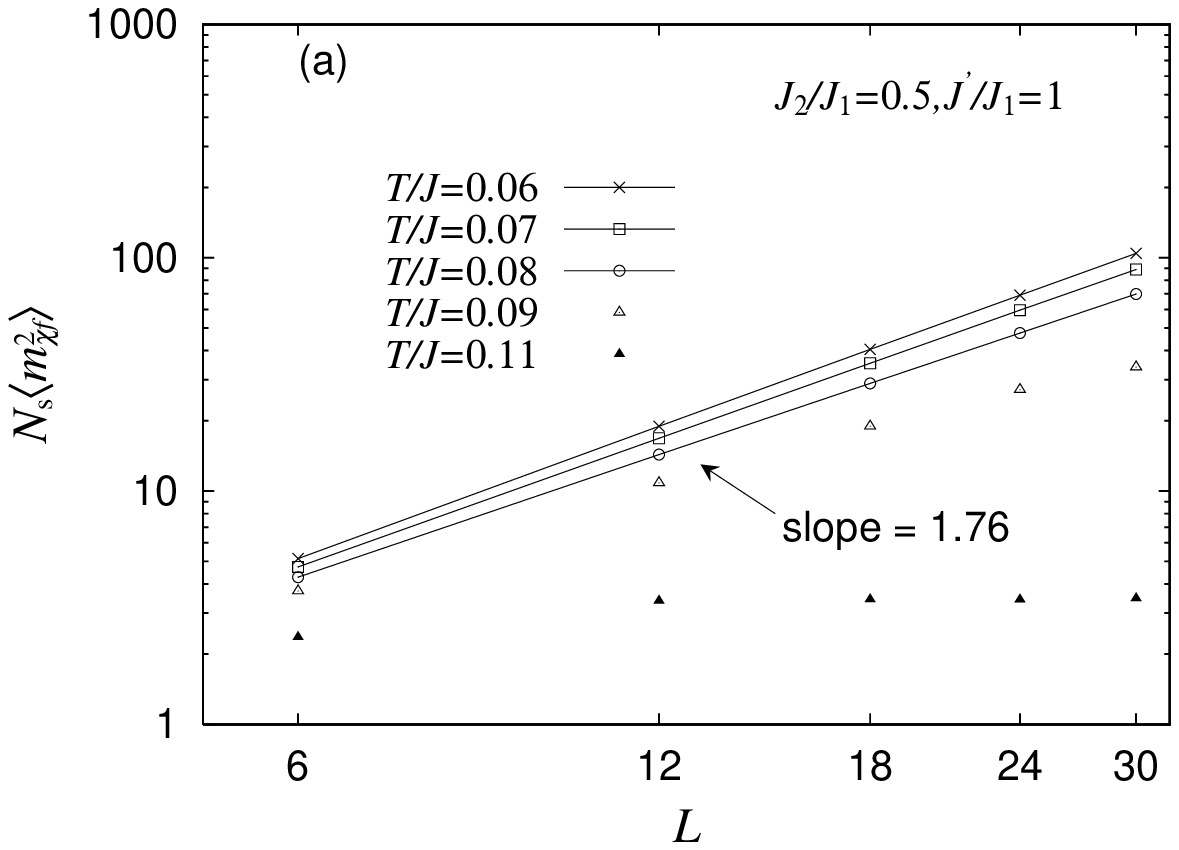}
\includegraphics[scale=0.8]{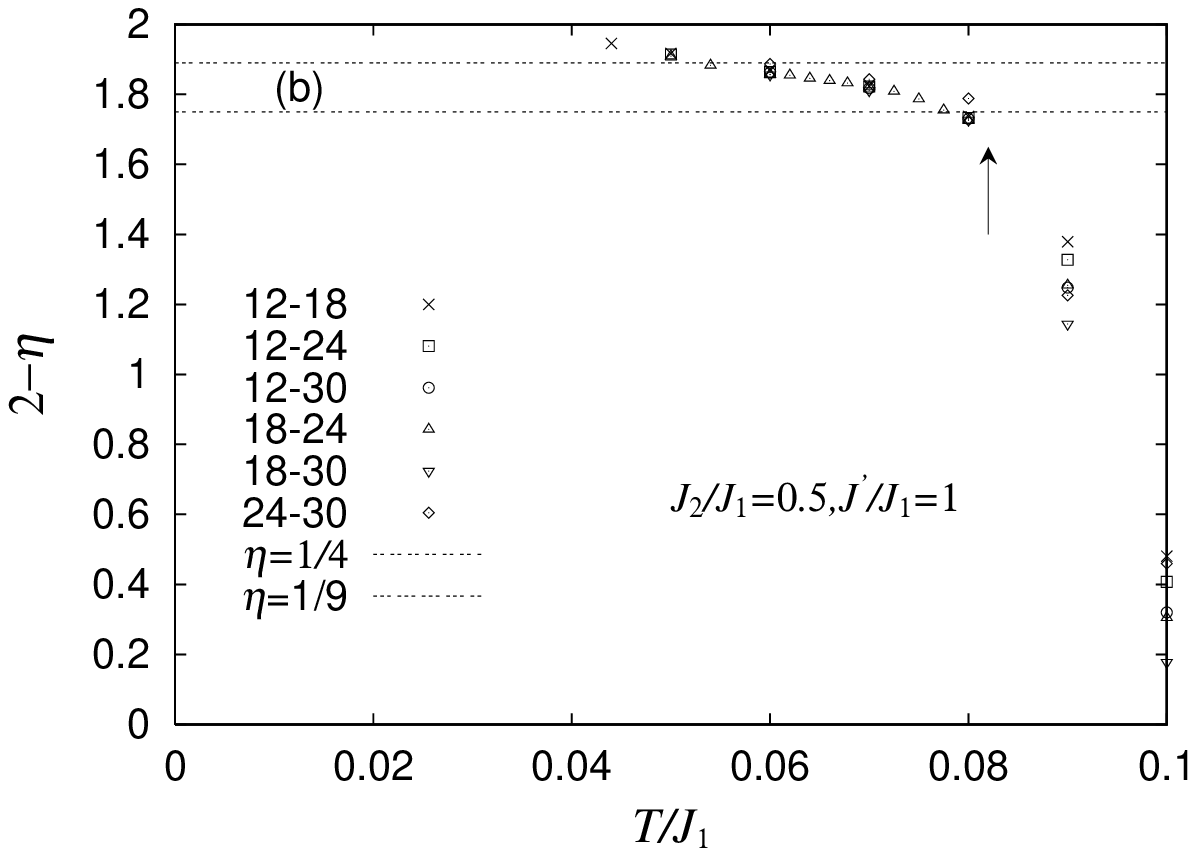}
\caption{(a) Log-log plot of the ferrimagnetic (staggered)
chiral susceptibility  versus the system size $L$ at
several temperatures, for the case $J_2/J_1=0.5$ and $J'/J_1=1$.
(b) The slope of the data in (a), which should be equal to
$2-\eta $, is plotted versus the temperature. Two numbers
in the legend represent the $L$ and $L'$ values used in the fit:
See text for details. The broken horizontal lines represent
the lines corresponding to $\eta =1/4$ and $\eta =1/9$.}
\end{center}
\end{figure}

\newpage

\begin{figure}
\begin{center}
\includegraphics[scale=0.8]{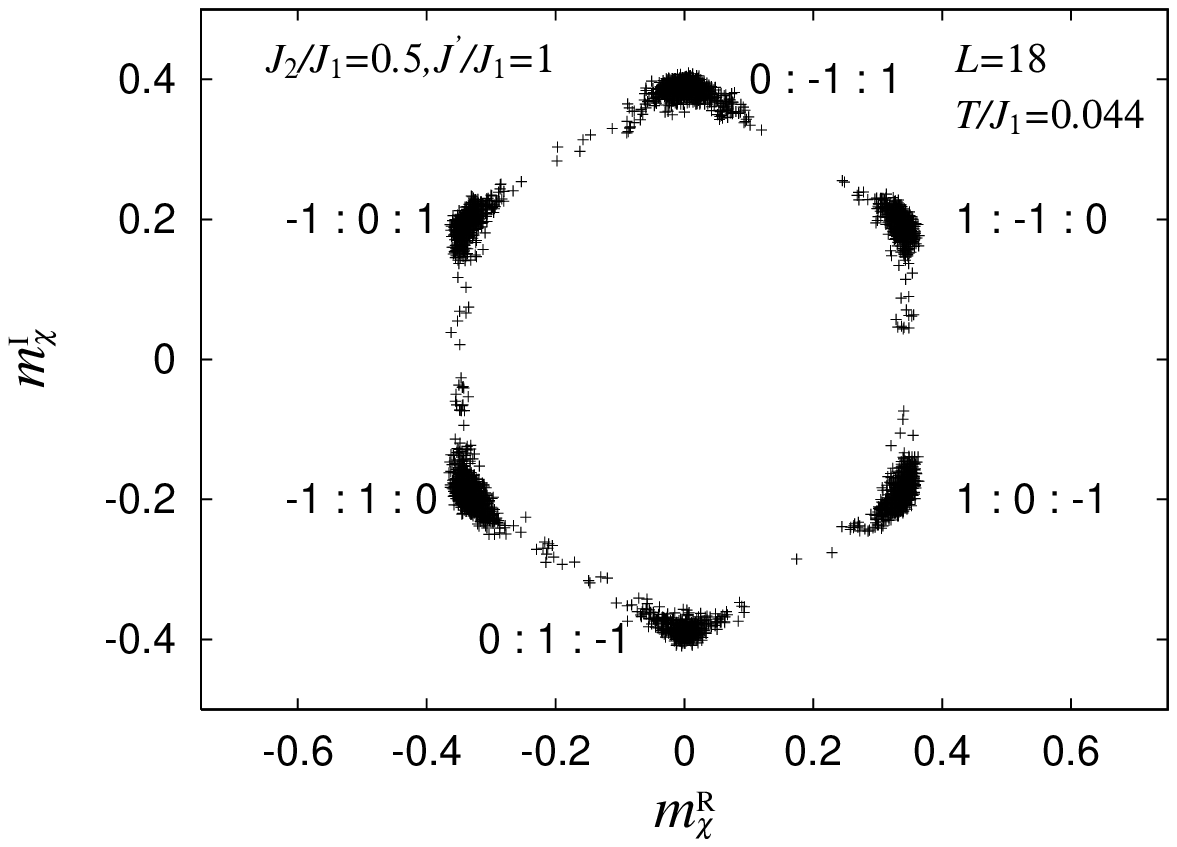}
\caption{Distribution of $(m_\chi^R,m_\chi^I)$, defined by eqs. (14) and (15),
for the case $J_2/J_1=0.5$ and $J'/J_1=1$.
The temperature is $T/J_1=0.044$, well below the transition
point $T_c/J_1\simeq 0.082$.
The lattice size is $L=18$.}
\end{center}
\end{figure}

\newpage

\begin{figure}
\begin{center}
\includegraphics[scale=0.8]{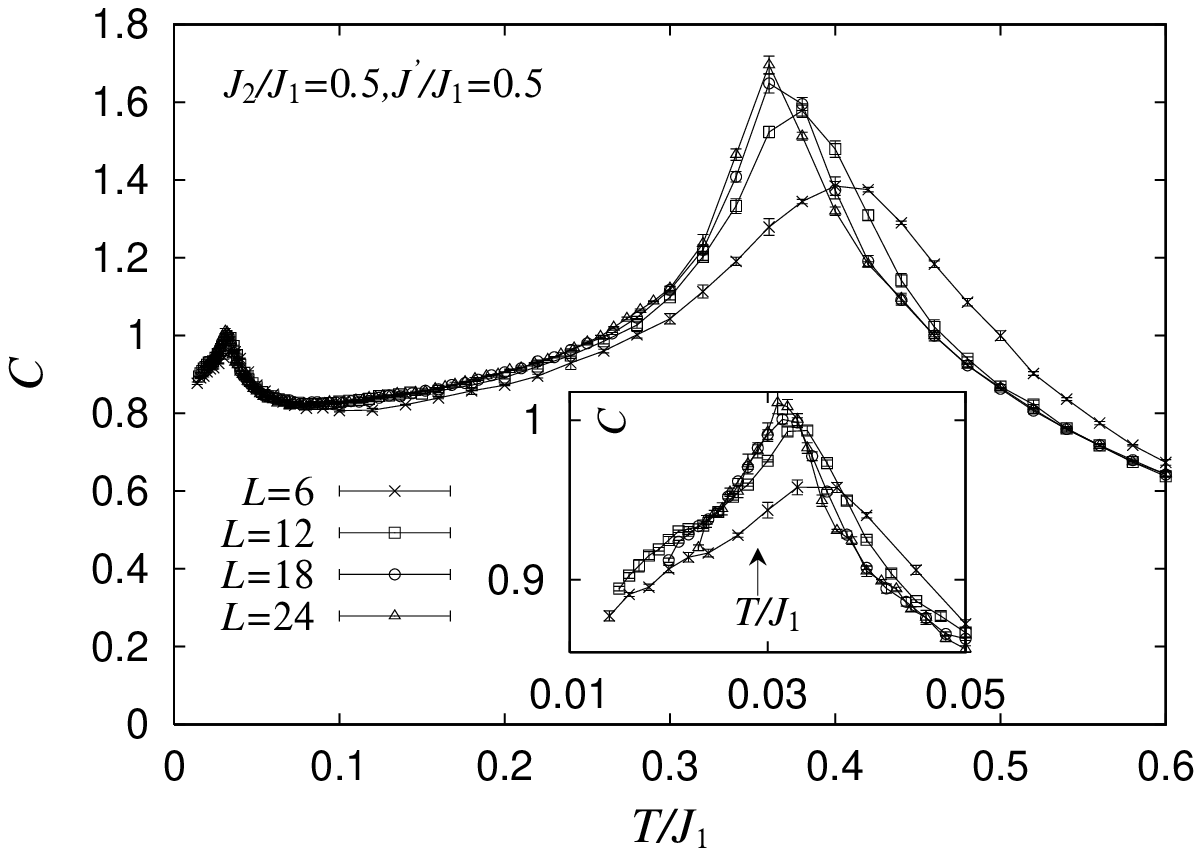}
\caption{The temperature and size dependence of the specific heat
per spin for the case $J_2/J_1=0.5$ and $J'/J_1=0.5$.  The inset is a
magnified view of the low-temperature regime. The arrow indicates
the transition point.}
\end{center}
\end{figure}

\begin{figure}
\begin{center}
\includegraphics[scale=0.8]{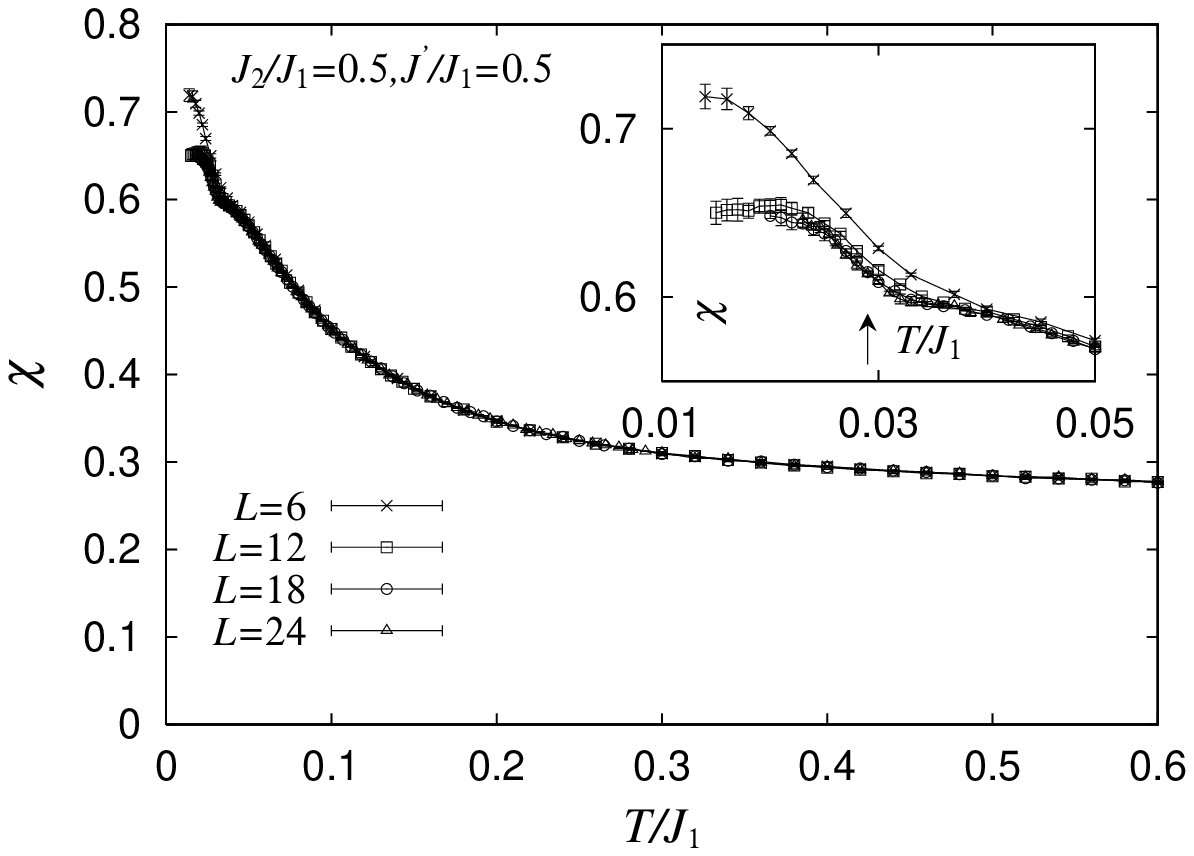}
\caption{The temperature and size dependence of the linear
susceptibility
per spin for the case $J_2/J_1=0.5$ and $J'/J_1=0.5$.
The inset is a
magnified view of the low-temperature regime.
The arrow indicates the transition point.}
\end{center}
\end{figure}

\newpage

\begin{figure}
\begin{center}
\includegraphics[scale=0.8]{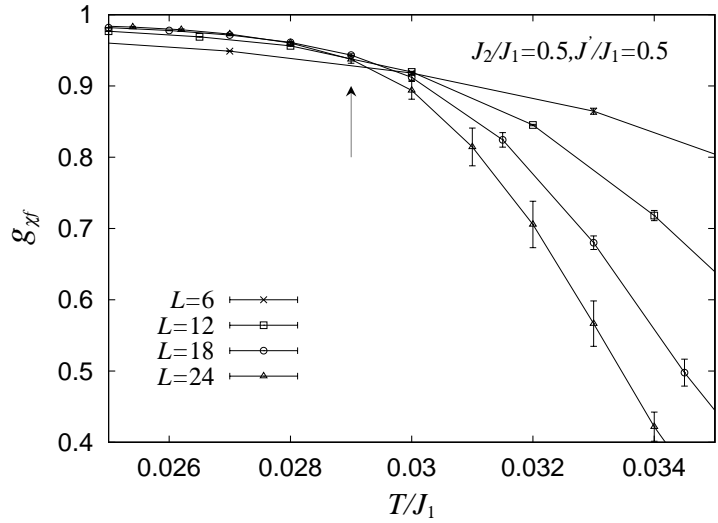}
\caption{The temperature and size dependence of the chiral
Binder ratio for the case $J_2/J_1=0.5$ and $J'/J_1=0.5$
in the transition region.
The arrow indicates the transition point.}
\end{center}
\end{figure}

\clearpage

\begin{figure}
\begin{center}
\includegraphics[scale=0.8]{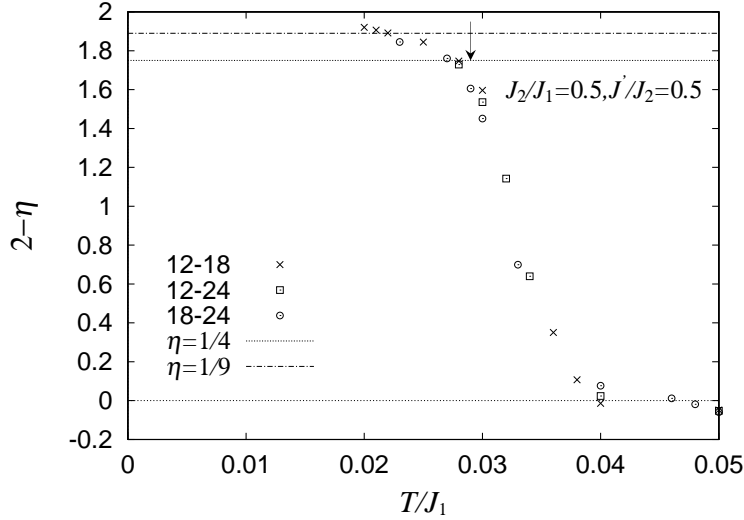}
\caption{The estimated exponent $2-\eta $ is plotted versus the
temperature for the case $J_2/J_1=0.5$ and $J'/J_1=0.5$.
The two numbers
in the legend represent the $L$ and $L'$ values used in the fit:
See text for details. The dashed horizontal lines represent
the lines corresponding to $\eta =1/4$ and $\eta =1/9$.}
\end{center}
\end{figure}

\newpage

\begin{figure}
\begin{center}
\includegraphics[scale=0.8]{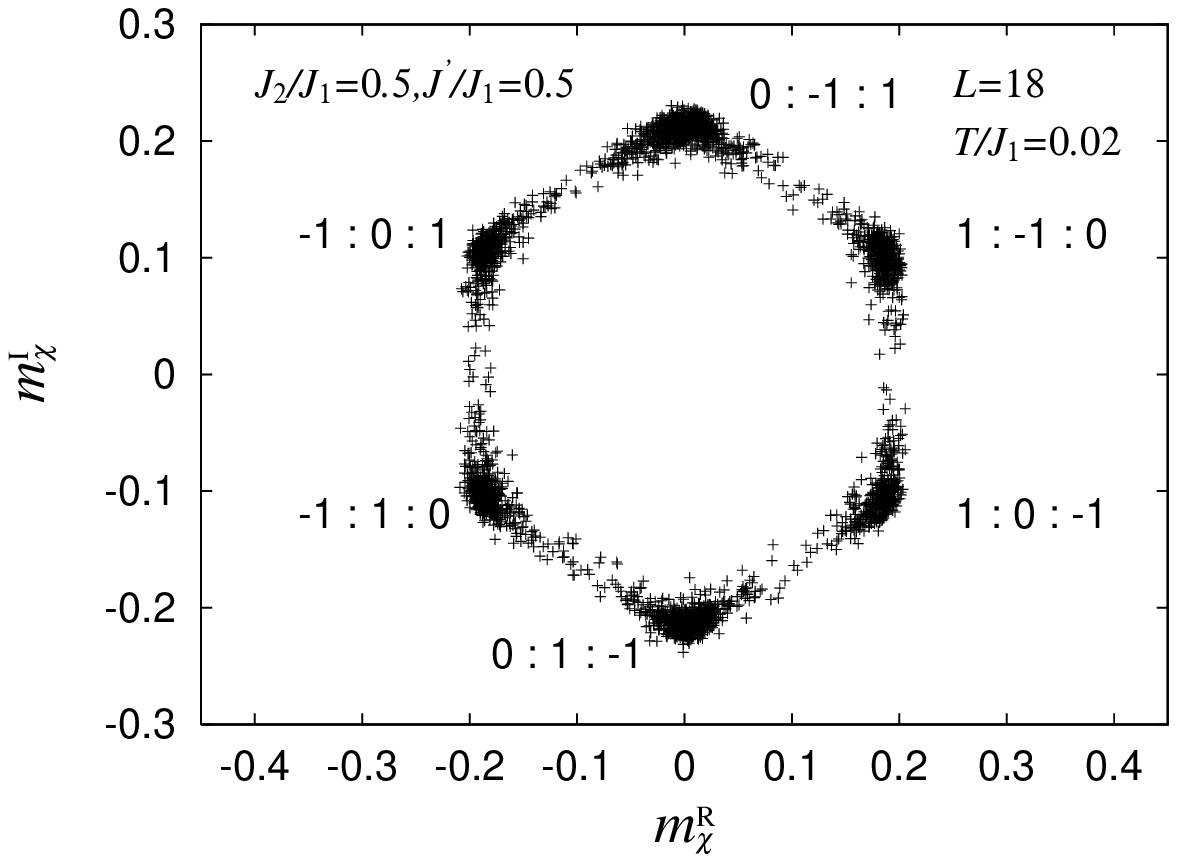}
\caption{Distribution of $(m_\chi^R,m_\chi^I)$, defined by eqs. (14) and
(15), for the case $J_2/J_1=0.5$ and $J'/J_1=0.5$.
The temperature is $T/J_1=0.02$, well below the transition
point $T_c/J_1\simeq 0.029$.
The lattice size is $L=18$.}
\end{center}
\end{figure}

\newpage

\begin{figure}
\begin{center}
\includegraphics[scale=0.8]{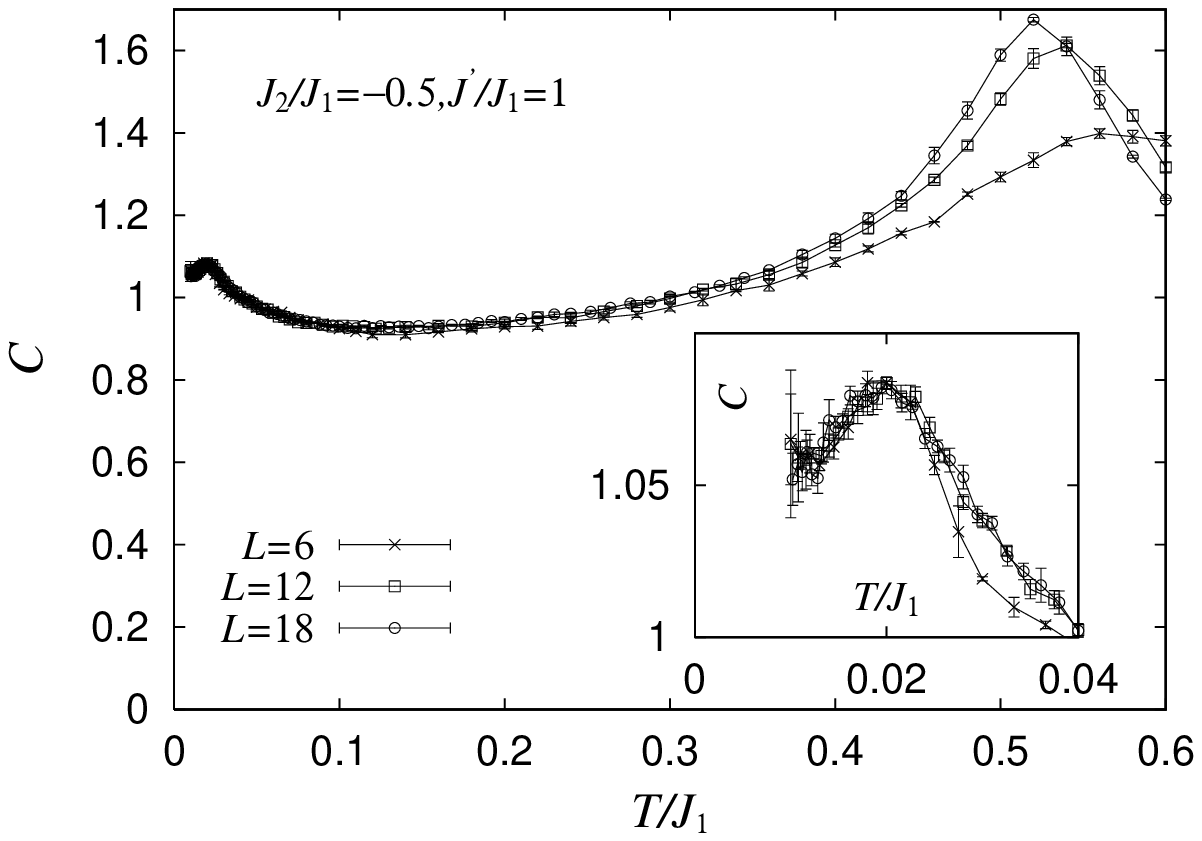}
\caption{The temperature and size dependence of the specific heat
per spin for the case $J_2/J_1=-0.5$ and $J'/J_1=1$. The inset is a
magnified view of the low-temperature regime.}
\end{center}
\end{figure}

\newpage

\begin{figure}
\begin{center}
\includegraphics[scale=0.8]{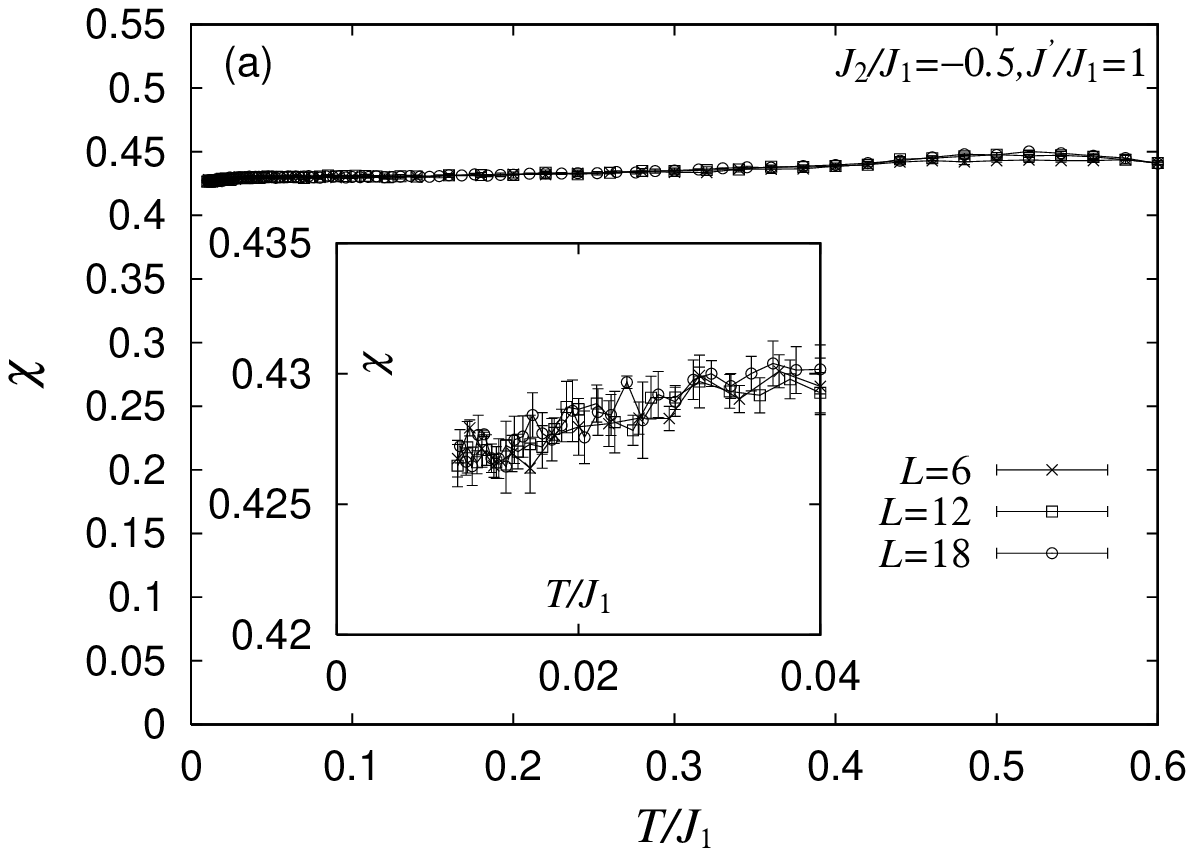}
\hspace*{1mm}
\includegraphics[scale=0.79]{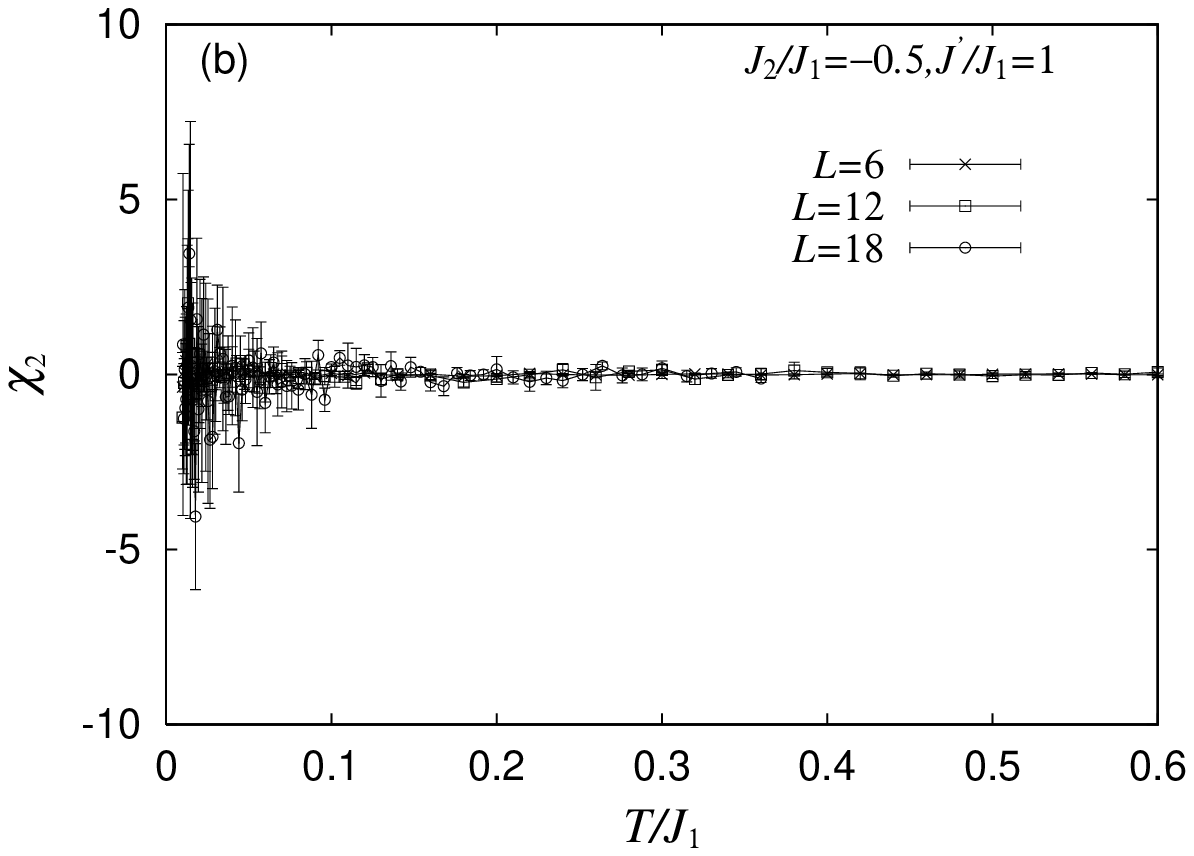}
\caption{The temperature and size dependence of (a) the linear
susceptibility per spin, and (b) the nonlinear susceptibility per spin, 
for the case $J_2/J_1=-0.5$ and $J'/J_1=1$.
The inset is a
magnified view of the low-temperature regime.}
\end{center}
\end{figure}

\newpage

\begin{figure}
\begin{center}
\includegraphics[scale=0.8]{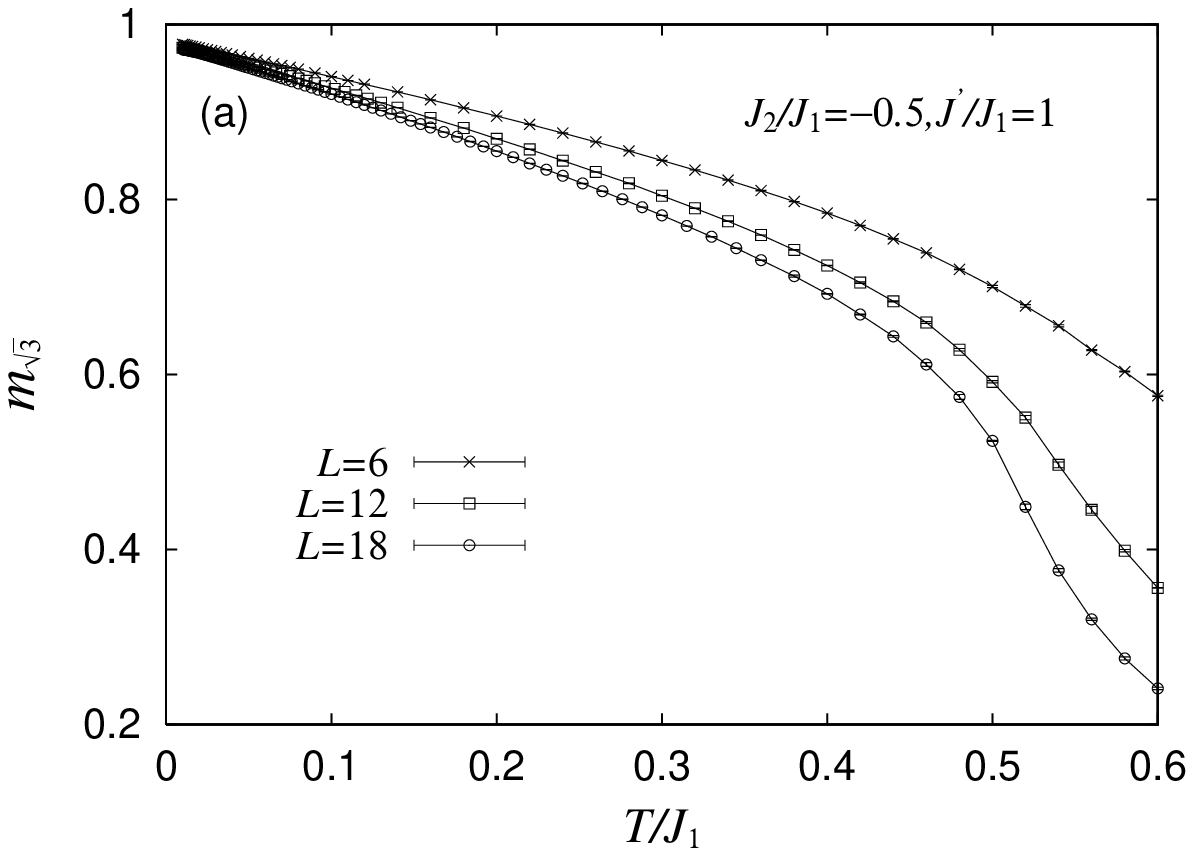}
\includegraphics[scale=0.8]{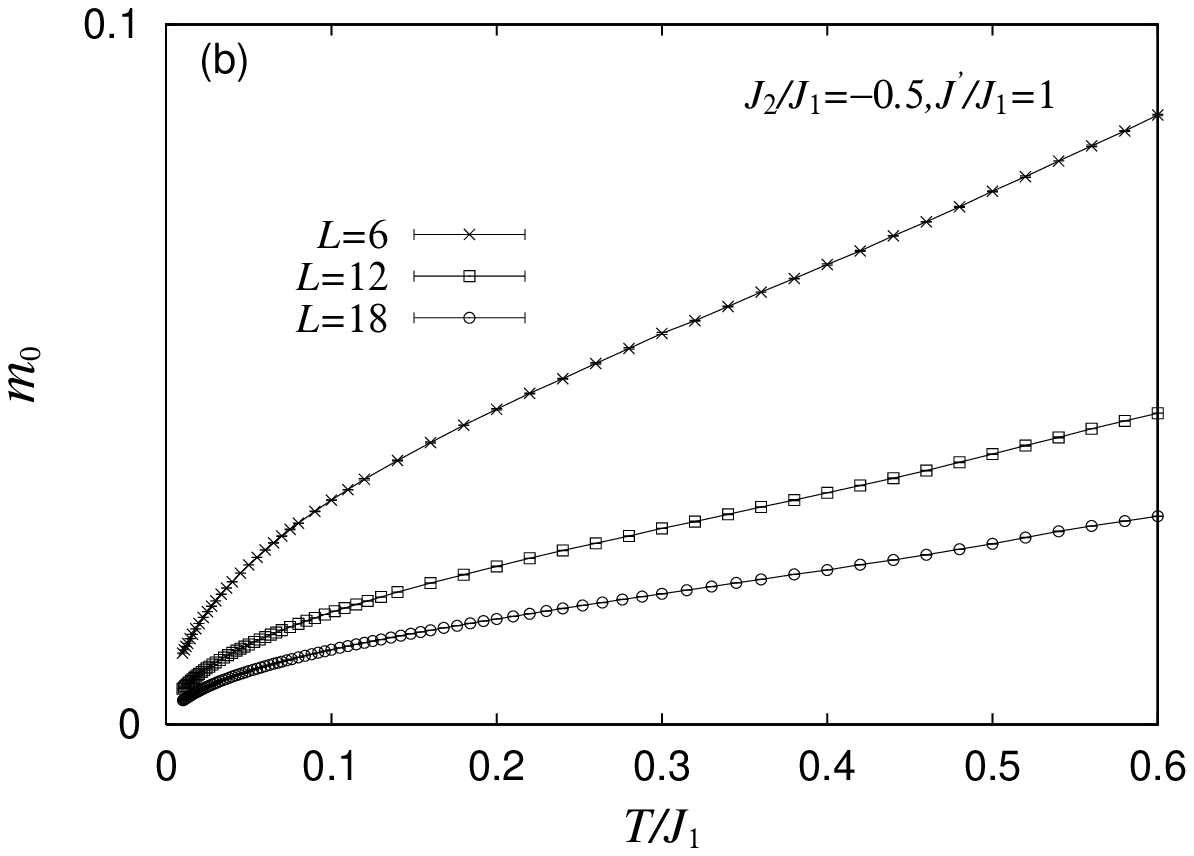}
\caption{The temperature and size dependence of the
ferrimagnetic magnetization per spin
associated with the (a)$\sqrt 3\times \sqrt 3$ mode, and 
(b) the $q=0$ mode, for the case $J_2/J_1=-0.5$ and $J'/J_1=1$.}
\end{center}
\end{figure}

\newpage

\begin{figure}
\begin{center}
\includegraphics[scale=0.8]{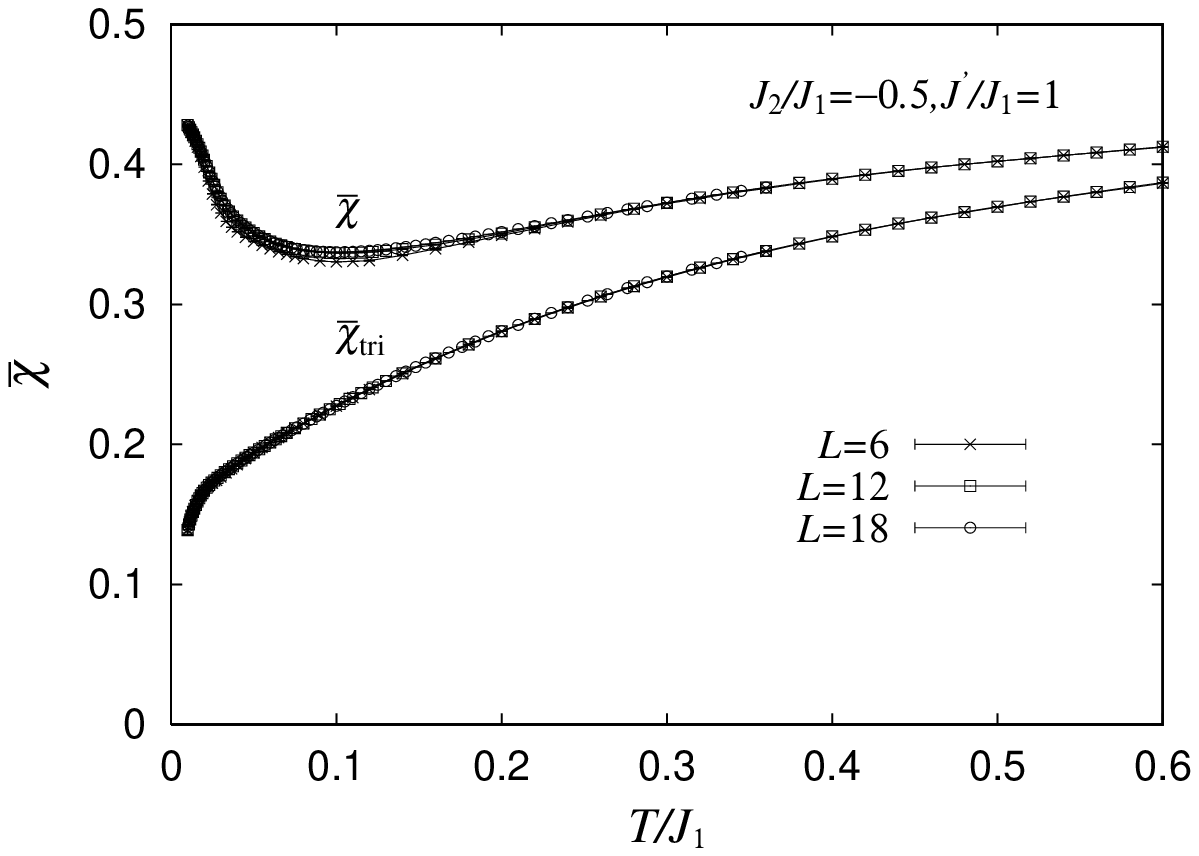}
\caption{The temperature and size dependence of the mean local
amplitude of the chirality
for the case $J_2/J_1=-0.5$ and $J'/J_1=1$. $\bar \chi$ and
$\bar \chi_{{\rm tri}}$ represent the local chirality defined at the
upward and downward triangles on the lower kagom\'e layer.}
\end{center}
\end{figure}

\begin{figure}
\begin{center}
\includegraphics[scale=0.8]{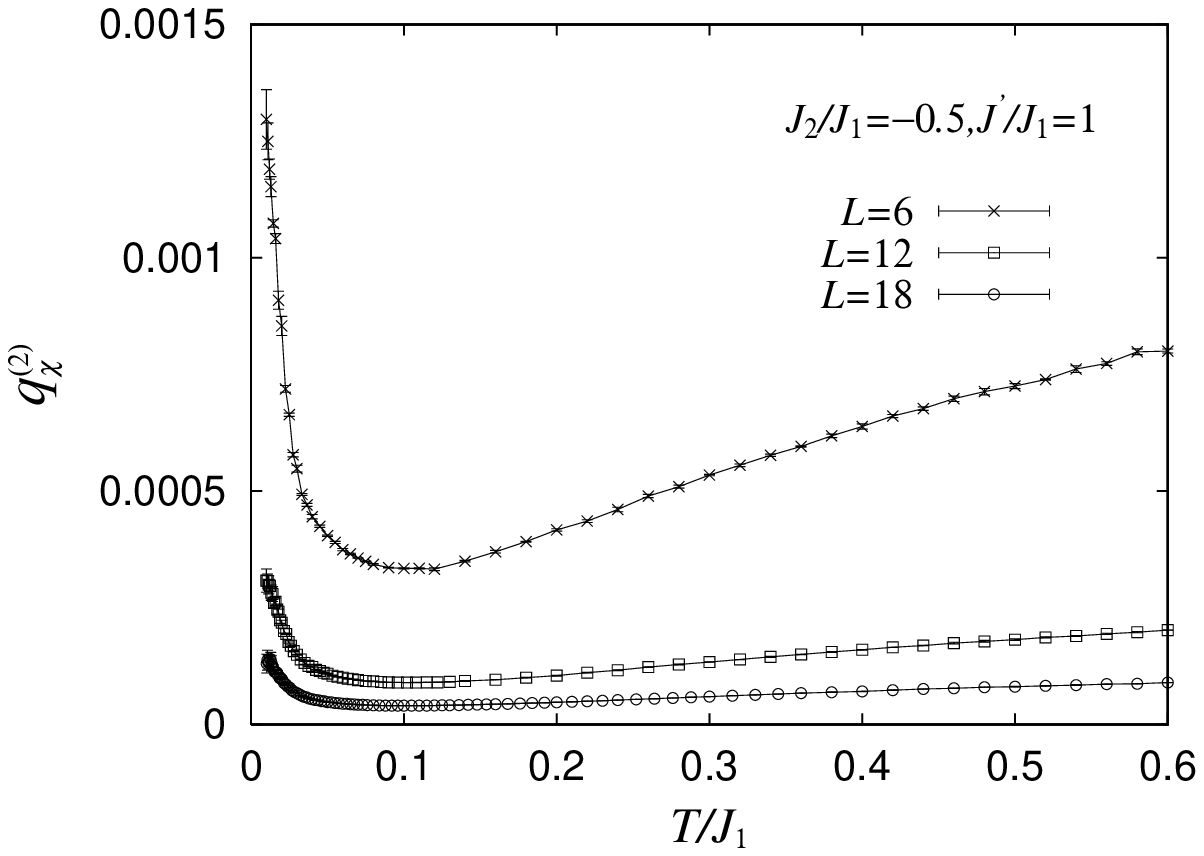}
\caption{The temperature and size dependence of the
chiral-glass order parameter
for the case $J_2/J_1=-0.5$ and $J'/J_1=1$.}
\end{center}
\end{figure}

\newpage

\begin{figure}
\begin{center}
\includegraphics[scale=0.8]{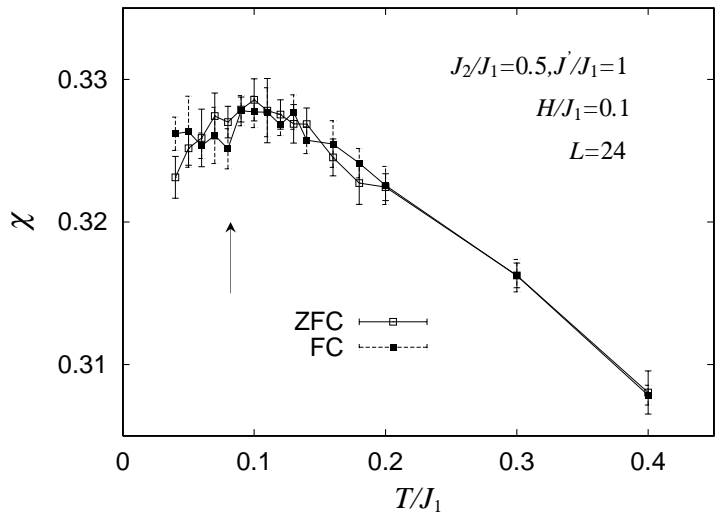}
\caption{The temperature dependence of the field-cooled (FC) and
zero-field-cooled (ZFC) susceptibilities for the case
$J_2/J_1=0.5$ and $J'/J_1=1$. The intensity of the applied
field is $H/J_1=0.1$, and
the lattice size is $L=24$.
At each temperature, total of
10$^6$ MCS are generated, of which the latter 5$\times10^5$ MCS 
are used to measure the magnetization.
FC magnetization is measured by gradually cooling the system
in a field
in steps from $T/J_1=0.4$ to $T/J_1=0.04$ across the
transition temperature $T_c/J_1\simeq 0.082$.
ZFC  magnetization is measured, first by quenching the
system from infinite temperature to $T/J_1=0.04$, and then
by gradually warming the system in a field
in steps from $T/J_1=0.04$ to $T/J_1=0.4$ across the
transition temperature.}
\end{center}
\end{figure}